\documentclass[fleqn,usenatbib]{mnras}

\usepackage{newtxtext,newtxmath}

\usepackage[T1]{fontenc}
\usepackage{graphicx}
\usepackage{amsmath}
\usepackage{amssymb}
\usepackage[T1]{fontenc}
\usepackage{ae,aecompl}
\usepackage{comment}
\usepackage{bm}
\usepackage{color}
\usepackage[normalem]{ulem}
\usepackage{hyperref}
\hypersetup{
    colorlinks=true,
    linkcolor=blue,
    filecolor=magenta,      
    urlcolor=cyan,
}

\numberwithin{equation}{section}

\definecolor{webgreen}{rgb}{0,.5,0}
\definecolor{webbrown}{rgb}{.6,0,0}

\newcommand{\be}{\begin{equation}}
\newcommand{\ee}{\end{equation}}
\newcommand{\ba}{\begin{eqnarray}}
\newcommand{\ea}{\end{eqnarray}}
\newcommand{\bse}{\begin{subequations}}
\newcommand{\ese}{\end{subequations}}

\usepackage{etoolbox}

\title[Cooling flows around CGM clouds]{Cooling flows around cold clouds in the circumgalactic medium: steady-state models \& comparison with TNG50}

\author[Alankar Dutta, Prateek Sharma, Dylan Nelson]{Alankar Dutta$^{1}$\thanks{{alankardutta@iisc.ac.in}},
Prateek Sharma$^{1}$\thanks{{prateek@iisc.ac.in}} and Dylan Nelson$^{2}$\thanks{{dnelson@uni-heidelberg.de}} \\
$^{1}$Department of Physics, Indian Institute 
of Science, Bangalore, India - 560012\\
$^{2}$Universit\"{a}t Heidelberg, Zentrum f\"{u}r Astronomie, Institut f\"{u}r theoretische Astrophysik, Albert-Ueberle-Str. 2, 69120 Heidelberg, Germany}

\date{}
\pubyear{2021}

\begin{document}
\label{firstpage}
\pagerange{\pageref{firstpage}--\pageref{lastpage}}
\maketitle

\begin{abstract}
Cold, non-self-gravitating clumps occur in various astrophysical systems, ranging from the interstellar and circumgalactic medium (CGM), to AGN outflows and solar coronal loops. Cold gas has diverse origins such as turbulent mixing or precipitation from hotter phases. We obtain the analytic solution for a steady pressure-driven 1-D cooling flow around cold, local over-densities, irrespective of their origin. Our solutions describe the slow and steady radiative cooling-driven gas inflow in the saturated regime of nonlinear thermal instability in clouds, sheets and filaments. Such a cooling flow develops when the gas around small clumps undergoes radiative cooling. These small-scale, cold `seeds' are embedded in a large volume-filling hot CGM maintained by feedback. 
We use a simple two-fluid treatment to include magnetic fields as an additional polytropic fluid. To test the limits of applicability of these analytic solutions, we compare with the gas structure found in and around small-scale cold clouds in the CGM of massive halos in the TNG50 cosmological MHD simulation from the IllustrisTNG suite. Despite qualitative resemblance of the gas structure, we find deviations from steady state profiles generated by our model. Complex geometries and turbulence all add complexity beyond our analytic solutions. We derive an exact relation between the mass cooling rate ($\dot{\rm M}_{\rm cool}$) and the radiative cooling rate ($\dot{\rm E}_{\rm cool}$) for a steady cooling flow. A comparison with the TNG50 clouds shows that this cooling flow relation only applies in a narrow temperature range around $\rm \sim 10^{4.5}$\,K where the isobaric cooling time is the shortest. In general, turbulence and mixing, instead of radiative cooling, may dominate the transition of gas between different temperature phases.
\end{abstract}

\begin{keywords}
galaxies: evolution -- clusters: intracluster medium -- ISM: clouds -- galaxies: haloes
\end{keywords}


\section{Introduction}\label{sec:introduction}

Multiphase plasmas are ubiquitous, occurring across a range of mass and length scales, from magnetic loops in the lower solar corona (\citealt{Reale1996,Kleint2014}), to flows around supermassive black holes (\citealt{Tremblay2016}), and in the circumgalactic (for a recent review, see \citealt{Tumlinson2017}) and intracluster medium (\citealt{McDonald2010,Voit2015}). Star-forming molecular clouds may condense out of, and grow at the expense of, the surrounding thermally unstable gas, with only the densest clumps becoming gravitationally unstable to collapse (\citealt{Wareing2019}).

The origins of multiphase gas across these diverse systems can be broadly classified into two categories: (i) the spontaneous condensation of cold gas from a hotter phase, if the ratio of the cooling time and the dynamical time is sufficiently small (\citealt{Sharma2012,Choudhury2019,Xia2017,Klimchuk2019}); (ii) the entrainment of mass on to a sufficiently large pre-existing cold gas cloud from the surrounding diffuse gas \citep{Armillotta2016}, due to mixing-driven radiative cooling in a boundary layer \citep{Gronke2018,Ji2019,Fielding2020}.

Cold gas structures are frequently observed to be surrounded by warm, intermediate-temperature gas with a short cooling time (\citealt{Schrijver2001,Fabian2003}), which can cool and accrete on to the cold seeds. This phenomenon has also been seen in numerical simulations, from individual clouds and multiphase winds to galactic halos (\citealt{Sharma2010,Vijayan2018,Waters2019a,Nelson2020,Schneider2020,Das2021}). 

In this paper we present one-dimensional pressure-driven steady cooling flow solutions in spherical, cylindrical and cartesian geometries. These can describe the local flows on to cold clumps in a multiphase medium. We use the words `clouds' and `clumps' interchangeably, and generally study the gas dynamics around cold gas structures. We generalize the hydrodynamic cooling flow solution to include magnetic fields as a polytropic fluid. We compare these solutions with the properties of cool/dense gas around clouds in the circumgalactic medium (CGM) of a $\sim 10^{13} {\rm M}_\odot$ halo in the TNG50 cosmological galaxy formation simulation.

We start with our analytic model and its solutions in section \ref{sec:PDCF}, including the effects of magnetic fields. In section \ref{sec:num_sims} we test our analytical results against 1-D hydrodynamical calculations. In section \ref{sec:tng} we then compare with local flows around cool clouds in the CGM of TNG50 halos. Section \ref{sec:discussion} discusses the astrophysical implications and the general applicability of our solutions. Section \ref{sec:conclusions} concludes and summarizes the key results.


\section{Pressure-driven cooling flow} \label{sec:PDCF}

The flow of a radiatively cooling gas on to an over-dense and lower pressure region can be described by a steady cooling flow solution. 
Consider a one-dimensional solution in cartesian, cylindrical and spherical geometries with the ideal gas equation of state. The mass, momentum and entropy equations in steady state are then
\bse
\label{eq:simplHydro}
\begin{align}
\label{eq:Mdot}
\dot{\rm M} \rm & = \rm - K  r^q \rho v,\\ 
\label{eq:mom}
\rm v \frac{dv}{dr} & = \rm - \frac{1}{\rho}\frac{dp}{dr}, \\
\label{eq:ent}
\rm \frac{pv}{(\gamma-1)} \frac{d}{dr}\left[ \ln \left( \frac{p}{\rho^\gamma} \right) \right]& = \rm - n_e n_i \Lambda(T),
\end{align}
\ese
where $\dot{\rm M}$ is the constant mass inflow rate, $\rm q=\{0,1,2\}$ and $\rm K=\rm\{ A,2\pi H,4\pi\}$ for cartesian, cylindrical and spherical geometries of the flow, respectively ($\rm A$ is the transverse area in cartesian geometry; $\rm H$ is the height of the cylinder). Here $\rm r$ is the coordinate distance, $\rm v$ ($\rm <0$ denotes inflowing) is the fluid velocity along this coordinate, and $\Lambda(T)$ is the temperature-dependent cooling function. We neglect any effects of self-gravity or external gravity in our equations, and implement the optically thin radiative cooling relevant for a plasma in collisional ionization equilibrium (e.g. \citealt{Sutherland1993}). For our cooling function, we use a Cloudy-generated (\citealt{2017RMxAA..53..385F}) cooling table with solar metallicity (mass fractions $X=0.7154$, $Y=0.2703$ and $Z=0.0142$ are taken from \citealt{2009ARA&A..47..481A}, temperatures going down to 10 K) in both our steady state ODE solution and in the time-dependent PDE solution presented in section \ref{sec:num_sims}.

The preceding system of three equations (Eqs. \ref{eq:simplHydro}) involves derivatives of three quantities (density, velocity, pressure) which can be used to obtain an equation containing only one derivative. Therefore, using Eqs. \ref{eq:Mdot} \& \ref{eq:ent}, Eq. \ref{eq:mom} can be written in the standard 
wind/accretion form,
\be
\label{eq:wind}
\rm \left( 1 - \frac{c_s^2}{v^2} \right) v \frac{dv}{dr} = \rm (\gamma-1)\frac{n_e n_i \Lambda}{\rho v} + \frac{q c_s^2}{r},
\ee
where $\rm c_s \equiv \sqrt{\gamma p/\rho}$ is the local sound speed. For $\rm q \ne 0$, this equation admits a critical point $\rm r_0$ where the right hand side vanishes.\footnote{Note that, unlike Parker wind or Bondi accretion solutions which are identical except for $\rm v \rightarrow -v$, here a critical point is possible only for inflow ($\rm v <0$). This is because the RHS of the cooling flow wind equation (Eq. \ref{eq:wind}), unlike the other two cases, has a velocity dependent term. Therefore, the direction (sign) matters if the RHS must vanish. In fact, an exclusively outflowing solution with a critical point is possible with a net heating instead of cooling.} There are two kinds of solutions with critical points, defined as the radius at which the right hand side of Eq. \ref{eq:wind} vanishes: (i) a transonic solution for which $\rm v(r_0) = \rm -c_s(r_0) = \rm -c_{s0}$ and the flow transitions from subsonic to supersonic as one crosses the critical point inwards; (ii) a fully subsonic or supersonic solution for which $\rm dv/dr$ vanishes at the critical point.\footnote{This second case is however physically unrealistic as we do not expect to find gas at supersonic speeds far from cooling sources. On the other hand, a transonic solution occurs only if the slope of the cooling function at the sonic point satisfies the condition derived in section \ref{sec:transonic_condn}. Finally, the critical point is also a sonic point for the transonic solution, but not for the subsonic solution.} At the critical point,
\be
\label{eq:r0_hydro}
\rm r_0= \rm q \,\gamma \,v_0 \,t_{\rm cool,0},
\ee
i.e., the advection and cooling times are comparable. Thus, the critical radius is larger for a higher advection velocity and a longer cooling time, and it can be much smaller than $\rm c_{s0} t_{\rm cool,0}$ for a subsonic flow. For the transonic solution, the additional requirement of $\rm v(r_0)= \rm -v_0= \rm -c_{s0}$ holds at the sonic point (note that we choose $\rm v_0$ to be positive and the cooling-flow velocity to be negative). Thus, the size of cold clumps, taken to be the sonic radius, is $\sim \rm c_{s0} t_{\rm cool,0}$, the only lengthscale in the problem (for the significance of this scale, see e.g., \citealt{Burkert2000,McCourt2018}). 

We can simplify our analysis and make the equations dimensionless by normalizing each variable with its value at the critical point, $\rm \tilde{r} = \rm r/r_0$, $\rm \tilde{\rho}= \rm \rho/\rho_0$, $\rm \tilde{v}= \rm v/v_0$, $\rm \tilde{p}= \rm p/p_0$, and $\rm \tilde{\Lambda}=\rm \Lambda(T)/\Lambda(T_0)$. As above, quantities subscripted by $\rm 0$ denote values evaluated at the critical point $\rm r_0$. In this case the de-dimensionalized equations become
\bse
\begin{align}
\label{eq:dd_Mdot}
1  \equiv \frac{\dot{\rm M}}{\rm K r_0^q \rho_0 v_{0}} & = \rm - \tilde{r}^q \tilde{\rho} \tilde{v},\\ 
\label{eq:dd_mom}
\rm \tilde{v} \frac{d\tilde{v}}{d\tilde{r}} & = \rm - \frac{1}{\gamma \tilde{\rho}}\frac{c_{s0}^2}{v_0^2}\frac{d\tilde{p}}{d\tilde{r}}, \\
\label{eq:dd_ent}
\rm \tilde{p}\tilde{v}\frac{d}{d\tilde{r}}\left[ \ln \left( \frac{\tilde{p}}{\tilde{\rho}^\gamma} \right)\right] & = \rm - q \gamma \tilde{\rho}^2 \tilde{\Lambda}, \\
\label{eq:dd_wind}
\rm \left( 1 - \frac{c_{s0}^2}{v_0^2}\frac{\tilde{p}}{\tilde{\rho}\tilde{v}^2} \right) \tilde{v} \frac{d\tilde{v}}{d\tilde{r}} & = \rm q \frac{c_{s0}^2}{v_0^2} \left[\frac{\tilde{\rho} \tilde{\Lambda}}{\tilde{v}} + \frac{\tilde{p}}{\tilde{r}\tilde{\rho}}\right].
\end{align}
\ese
At the critical point, $\rm T_0=(\mu m_p/ k_B) (p_0/\rho _0)$, where $\rm \mu$ is the mean particle mass in units of the proton mass. We note that by fixing these normalizations at the sonic point, the constant mass flux is automatically fixed to $\rm K r_0^q \rho_0 v_{0}$. In addition, our three first order ODEs require three boundary conditions, which we take as $\rm \rho_0$, $\rm v_0$ and $\rm c_{s0}$ at $ \rm r_0$ (which, in turn, is determined by the same parameters; see Eq. \ref{eq:r0_hydro}). 

We set up a convenient system of ODEs with two dependent variables $\rm \tilde{v}$ and $\rm \tilde{s} \equiv \tilde{p}/\tilde{\rho}^\gamma$,
\be
\label{eq:matrix_eqs}
\rm \frac{d}{d\tilde{r}} \begin{bmatrix} \rm \tilde{v} \\ \rm \tilde{s} \end{bmatrix} = \begin{bmatrix} \rm q \frac{c_{s0}^2}{v_0^2} \left(\frac{\tilde{\rho} \tilde{\Lambda}}{\tilde{v}} + \frac{\tilde{p}}{\tilde{r}\tilde{\rho}}\right) \big{/}
\left(\left[ 1 - \frac{c_{s0}^2}{v_0^2}\frac{\tilde{p}}{\tilde{\rho}\tilde{v}^2} \right] \tilde{v}\right) \\ \rm -q \gamma  \tilde{\Lambda} \tilde{\rho}^{(2-\gamma)}/\tilde{v} \end{bmatrix}.
\ee
The boundary condition at $\rm r=r_0$ ($\rm \tilde{r}=1$) is $\rm \tilde{v}=-1$, $\rm \tilde{\rho}=1$, $\rm \tilde{p}=1$, $\rm \tilde{s}=1$. We solve these equations moving outward and inward from the critical/sonic point. In analogy with the wind/accretion solutions, this system also admits both transonic and non-transonic (i.e. subsonic or supersonic throughout) solutions which may occur for appropriate boundary conditions.\footnote{Note that the cartesian cooling flow ($\rm q=0$) does not admit a critical point (i.e., neither the left nor right hand side of Eq. \ref{eq:wind} vanishes at any point) because the wind equation \ref{eq:wind} has a right hand side which can never be zero for $\rm q=0$. Moreover, in cartesian geometry, the Euler equations give $\rm p+\rho v^2$ to be a constant.}

Our equations represent a steady cooling flow driven by cooling and the associated pressure gradient, rather than by gravity as is more typically considered (\citealt{Stern2019}). These solutions can potentially describe the steady flows associated with the saturated state of nonlinear thermal instability in which cooling gas from the hot phase flows slowly on to cold and dense filaments/clouds (\citealt{Sharma2010}). Even the growth of cold seeds due to cooling of the mixed gas in the cloud-crushing problem (\citealt{Gronke2018,Waters2019b}) can be described qualitatively by these solutions, although a treatment for turbulent transport may be needed to adequately address this scenario.

\begin{figure}
	\includegraphics[width=\columnwidth]{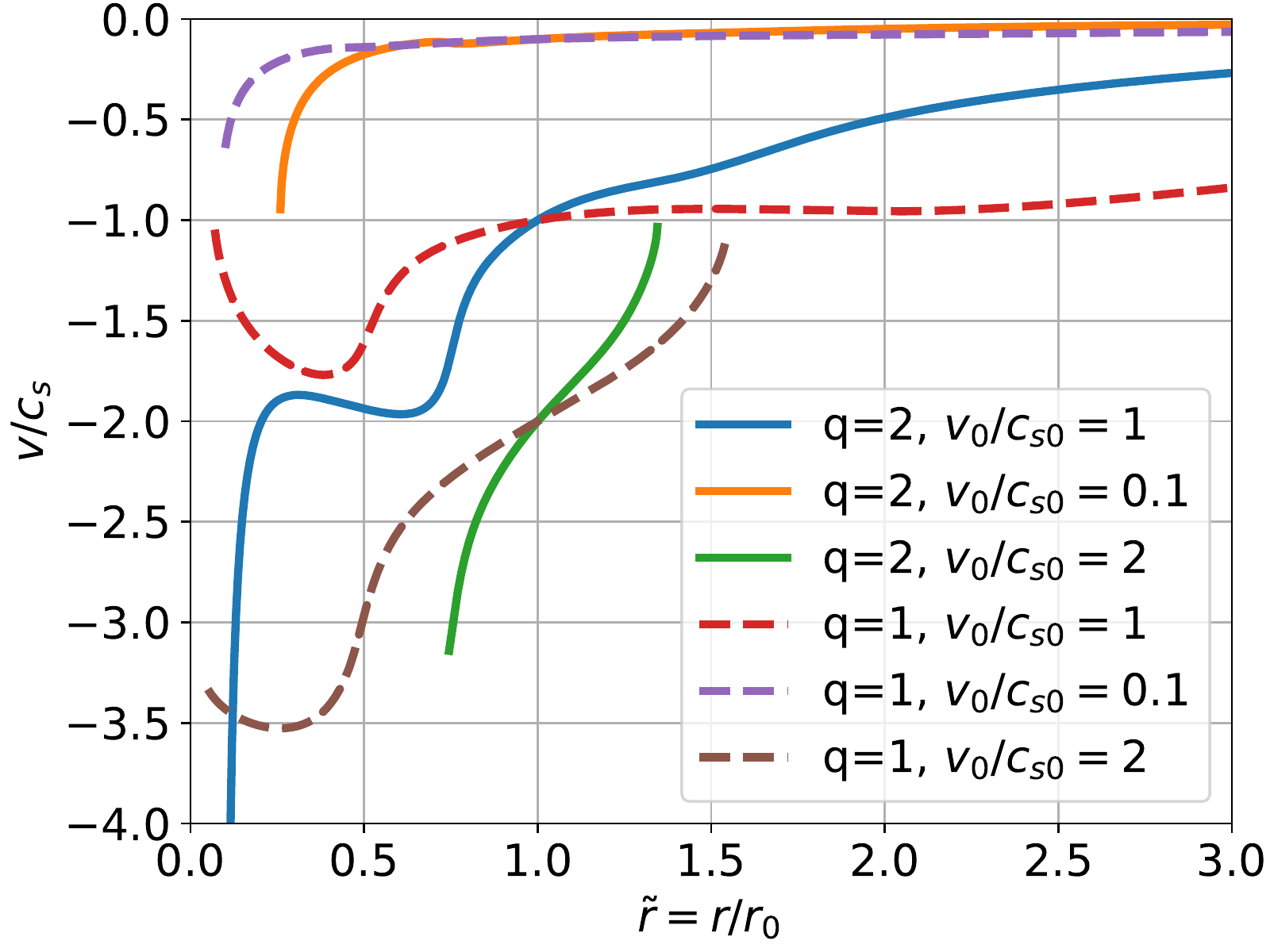}
	\caption [Mach vs r]{Some representative transonic, subsonic and supersonic solutions for spherical ($\rm q=2$) and cylindrical ($\rm q=1$) flow geometries. The critical temperature is $\rm T_0=4\times 10^4$ K and the critical density $\rm n_0=10^{-4}$ cm$^{-3}$. The shape of different profiles reflect the various features in the cooling curve. The profiles are truncated to a radial range in which we are able to obtain the solutions numerically (a steady-state solution is not present outside this as the numerical profiles become singular at the end points of this range). Note that $\rm v=c_s$ at $\rm \tilde{r}=1$ for the transonic solution, and $\rm dv/dr=0$ for the subsonic/supersonic solution (while a slope of $\rm v/c_s$ does not vanish at the critical point).}
	\label{fig:Machvsr}
\end{figure}

Figure \ref{fig:Machvsr} shows representative profiles of the Mach number as a function of the dimensionless radius for cooling flows in spherical (solid lines) and cylindrical (dashed lines) geometries. Subsonic, transonic and supersonic profiles are included. The nature of the solution depends on the shape of the cooling function. The cylindrical solutions are shallower than the spherical ones, and this trend is expected to continue to cartesian geometry that does not admit transonic solutions or solutions with extrema in velocity.


\subsection{The transonic solution}
\label{sec:transonic_condn}

The first equation in Eq. \ref{eq:matrix_eqs} has a $\rm 0/0$ form at the sonic point, provided a transonic solution exists. We can obtain the limiting value of the velocity derivative here by using L\'{}H\^{o}pital's rule. Applying Eqs. \ref{eq:dd_Mdot}$-$\ref{eq:dd_mom} at the sonic point, where $\rm \tilde{r}=\tilde{\rho}=\tilde{p}=\tilde{\Lambda}=-\tilde{v}=1$ and $\rm v_0=c_{s0}$, the radial gradients at the sonic point (denoted by a prime) are related as $\rm \tilde{\rho}^\prime = \tilde{v}^\prime - q$, $\rm \tilde{p}^\prime = \gamma \tilde{v}^\prime$. Plugging these in to Eq. \ref{eq:dd_wind}, where $\rm d\tilde{v}/d\tilde{r}$ has a $\rm 0/0$ form, we obtain
$$
\rm \tilde{v}^\prime = \frac{q \left[ 2(\tilde{v}^\prime-q) + \tilde{\Lambda}^\prime + \tilde{v}^\prime - \gamma \tilde{v}^\prime + 1\right]}{ (\tilde{v}^\prime-q) -2 \tilde{v}^\prime - \gamma \tilde{v}^\prime }.
$$
\begin{figure}
	\includegraphics[width=0.95\columnwidth]{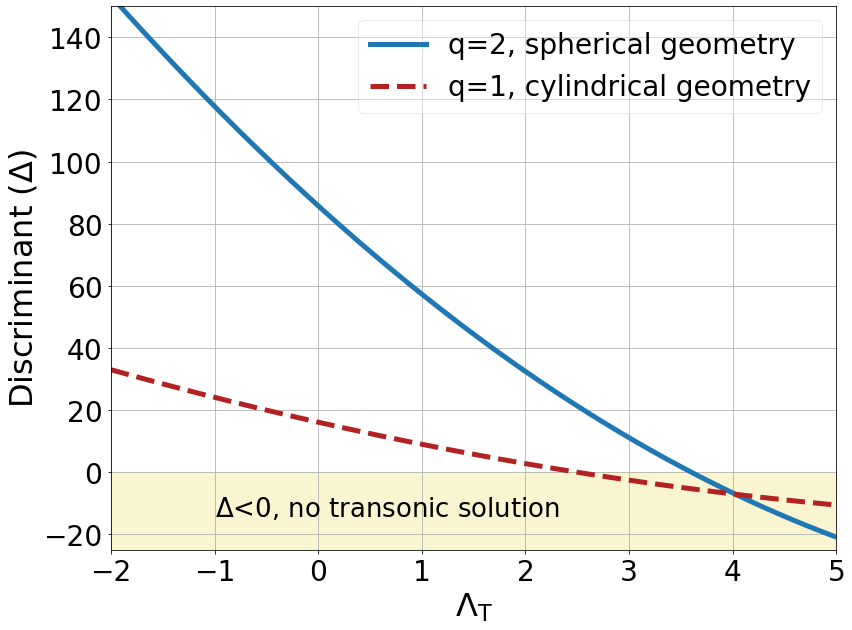}
	\includegraphics[width=0.95\columnwidth]{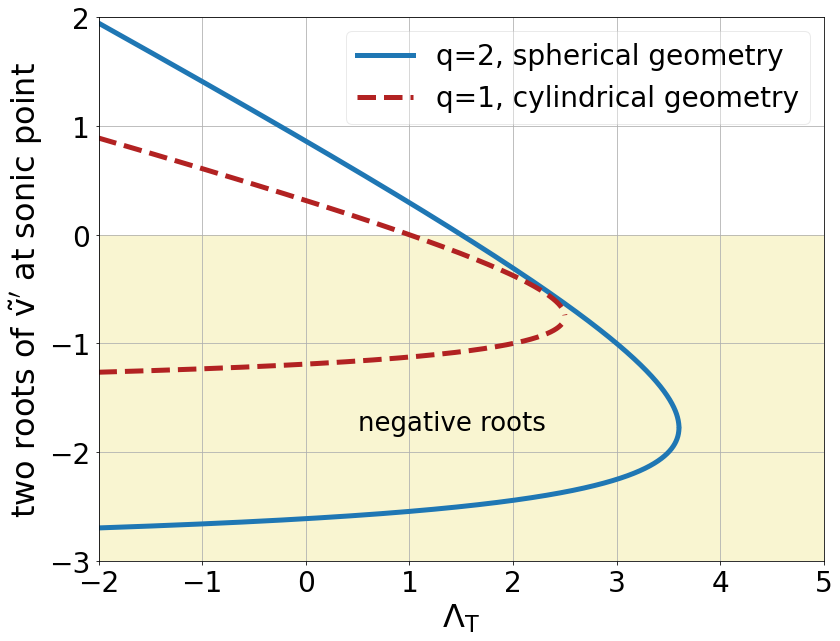}
	\caption[Discriminant]{\emph{Top panel:} The discriminant of the quadratic equation (Eq. \ref{eq:discriminant}) as a function of the slope of the cooling function at the sonic point ($\rm \Lambda_T$) for a steady spherical ($\rm q=2$) and cylindrical ($\rm q=1$) cooling flow. A transonic solution exists for $\rm \Lambda_T<3.6$ in spherical geometry and for $\rm \Lambda_T<2.5$ in cylindrical geometry. \emph{Bottom panel:} The two roots of the quadratic equation (Eq. \ref{eq:quadratic}) as a function of $\rm \Lambda_T$ for spherical and cylindrical geometries (when the discriminant is non-negative and allows real roots to exist). The positive root corresponds to a physically relevant transonic solution where fluid velocity decays to zero at large radii.}
	\label{fig:discriminant}
\end{figure}
\\
Expanding $\rm \tilde{\Lambda}^\prime$ as $\rm \tilde{\Lambda}^\prime = \Lambda_T (\tilde{p}^\prime - \tilde{\rho}^\prime) = \Lambda_T [(\gamma-1)\tilde{v}^\prime+q]$,  where $\rm \Lambda_T \equiv d\ln \Lambda/d\ln T$, we obtain the following quadratic equation for the velocity derivative at the sonic point,
\be
\label{eq:quadratic}
\rm (\gamma+1)\tilde{v}^{\prime 2} + q[\Lambda_T(\gamma-1)+4-\gamma] \tilde{v}^\prime + q[ q(\Lambda_T-2)+1 ] = 0.
\ee
The quadratic equation has a real solution only if the discriminant is non-negative; i.e., if
\be
\label{eq:discriminant}
\rm \Delta = q^2 [\Lambda_T(\gamma-1)+4-\gamma]^2 - 4(\gamma+1)q[q(\Lambda_T-2)+1] \geq 0.
\ee
The top panel of Figure~\ref{fig:discriminant} shows the discriminant, which is positive at all temperatures greater than $\rm 10^4$ K for the standard collisional ionization equilibrium cooling function (e.g., \citealt{Sutherland1993}), implying the existence of a transonic solution with appropriate boundary conditions. 

Although a positive discriminant ensures the existence of transonic solution, it may not be physically realizable. Physically relevant transonic solutions must have fluid velocity decaying to zero at large radii. This is possible if $\tilde{\rm v}^{\prime}$ is positive at the sonic point (as inflowing gas has a negative sign for velocity in our convention), meaning that only the real and positive roots of Eq.~\ref{eq:quadratic} are of physical interest. The bottom panel of Figure \ref{fig:discriminant} shows the values of $\rm \Lambda_T$ for which a positive $\tilde{\rm v}^\prime$ exists, which is possible for a sonic temperature larger than $10^4$ K for standard cooling functions.


\subsection{Including magnetic fields}

The circumgalactic medium is weakly magnetized, with the plasma $\rm \beta \equiv p_{\rm gas}/p_{\rm mag} \sim 100$ in the diffuse hot phase \citep{Nelson2020,Pakmor20}, meaning that the magnetic support in the hot phase is negligible. However, as the gas cools and compresses, the magnetic field can increase in the cooler phases because of flux freezing. As a result, cold gas phases in the CGM are expected to be magnetically dominated (\citealt{Sharma2010,Nelson2020}).

The magnetic pressure can be included in 1-D by modifying the momentum equation to
\be
\label{eq:mom_mag}
\rm v \frac{dv}{dr} = - \frac{1}{\rho}\frac{d}{dr} (p_{\rm gas} + p_{\rm mag}),
\ee
thereby including one more fluid component accounting for an additional magnetic pressure which follows a polytropic equation (c.f. Figure~\ref{fig:mag-indx}). This particular treatment of magnetic fields is mathematically similar to fluid models of adiabatic cosmic rays \citep[e.g.][]{Drury1981,Jun94}, and cosmic rays 
can also be included in an analogous manner.

The gas entropy evolution is still given by Eq. \ref{eq:dd_ent}. 
The magnetic pressure 
is assumed to satisfy a polytropic equation of state,
\be
\label{eq:p_mag}
\rm \frac{d}{dr}  \left( \frac{p_{\rm mag}}{\rho^{\gamma_m}} \right) = 0,
\ee
where $\rm \gamma_m$ is the polytropic index for magnetic pressure. This index depends on the magnetic and compression geometry ($\rm \gamma_m=4/3$ for isotropic conditions, 0 for gas compression along field lines, and 2 for compression across field lines), and is a consequence of flux-freezing. Note that this approach of including magnetic effects is only approximate and excludes effects such as the generation of magnetic fields due to turbulence.

The wind equation (Eq.~\ref{eq:wind}) in presence of magnetic fields becomes
\be
\label{eq:wind_mag}
\rm \left( 1 - \frac{c_t^2}{v^2} \right) v \frac{dv}{dr} = (\gamma-1)\frac{n_e n_i \Lambda}{\rho v} + \frac{q c_t^2}{r},
\ee
where we introduce the two-fluid sound speed $\rm c_t^2 = c_s^2 +c_m^2$, taking $\rm c_m \equiv \sqrt{\gamma_m p_{\rm mag}/\rho}$ as the magnetic signal speed. The de-dimensionalized momentum (Eq. \ref{eq:dd_mom}), wind (Eq. \ref{eq:dd_wind}), and magnetic pressure (Eq. \ref{eq:p_mag}) equations become
\bse
\begin{align}
\label{eq:dd_mom_mag}
&\rm \tilde{v} \frac{d\tilde{v}}{d\tilde{r}}  = - \frac{1}{\gamma \tilde{\rho}}\left(\frac{c_{s0}}{v_0}\right)^2 \left( \frac{d\tilde{p}_{\rm gas}}{d\tilde{r}} + \frac{1}{\beta_0}\frac{d\tilde{p}_{\rm mag}}{d\tilde{r}}\right), \\
\label{eq:dd_wind_mag}
\nonumber
&\rm \left[ 1 - \left(\frac{c_{s0}}{v_0}\right)^2 \left( \frac{\tilde{p}_{\rm gas}}{\tilde{\rho}\tilde{v}^2} \right) \left( 1 + \frac{\gamma_m}{\gamma \beta_0} \frac{\tilde{\rho}^\gamma}{\tilde{p}_{\rm gas}} \right) \right] \tilde{v} \frac{d\tilde{v}}{d\tilde{r}}  = \\
&\rm q \left(\frac{c_{s0}}{v_0}\right)^2 \left[\frac{\tilde{\rho} \tilde{\Lambda}}{\tilde{v}} \left( 1+ \frac{\gamma_m}{\gamma \beta_0}\right) + \frac{\tilde{p}_{\rm gas}}{\tilde{r}\tilde{\rho}} \left( 1 + \frac{\gamma_m}{\gamma \beta_0} \frac{\tilde{\rho}^{\gamma_m}}{\tilde{p}_{\rm gas}} \right) \right], \\
\label{eq:dd_pmag}
&\rm \frac{d}{d\tilde{r}} \left( \frac{\tilde{p}_{\rm mag}}{\tilde{\rho}^{\gamma_m}} \right) = 0
\end{align}
\ese
\\
where $\rm \beta_0 = p_{\rm gas,0}/p_{\rm mag,0}$ and $\rm \tilde{p}_{ mag}=p_{\rm mag}/p_{ mag,0}$. The inclusion of magnetic fields leads to the additional parameter $\rm \beta_0$, and for $\rm \beta_0 \rightarrow \infty$ we recover the pure hydro solution. Here $\rm p_{\rm mag,0}$ is the magnetic pressure at the critical point, which is modified in the presence of magnetic fields to 
\be
\label{eq:r0_MHD}
\rm r_0 = q \,\gamma \,v_0 \,t_{\rm cool,0} \left ( 1+ \frac{\gamma_m}{\gamma \beta_0} \right).
\ee
As expected, the additional magnetic pressure pushes the critical point outwards. For a transonic solution, the velocity at the sonic point is now $\rm v_0 = c_{t0} = [(\gamma p_{\rm gas,0} + \gamma_m p_{\rm mag,0})/\rho_0]^{1/2}$. 

\begin{figure}
	\includegraphics[width=\columnwidth]{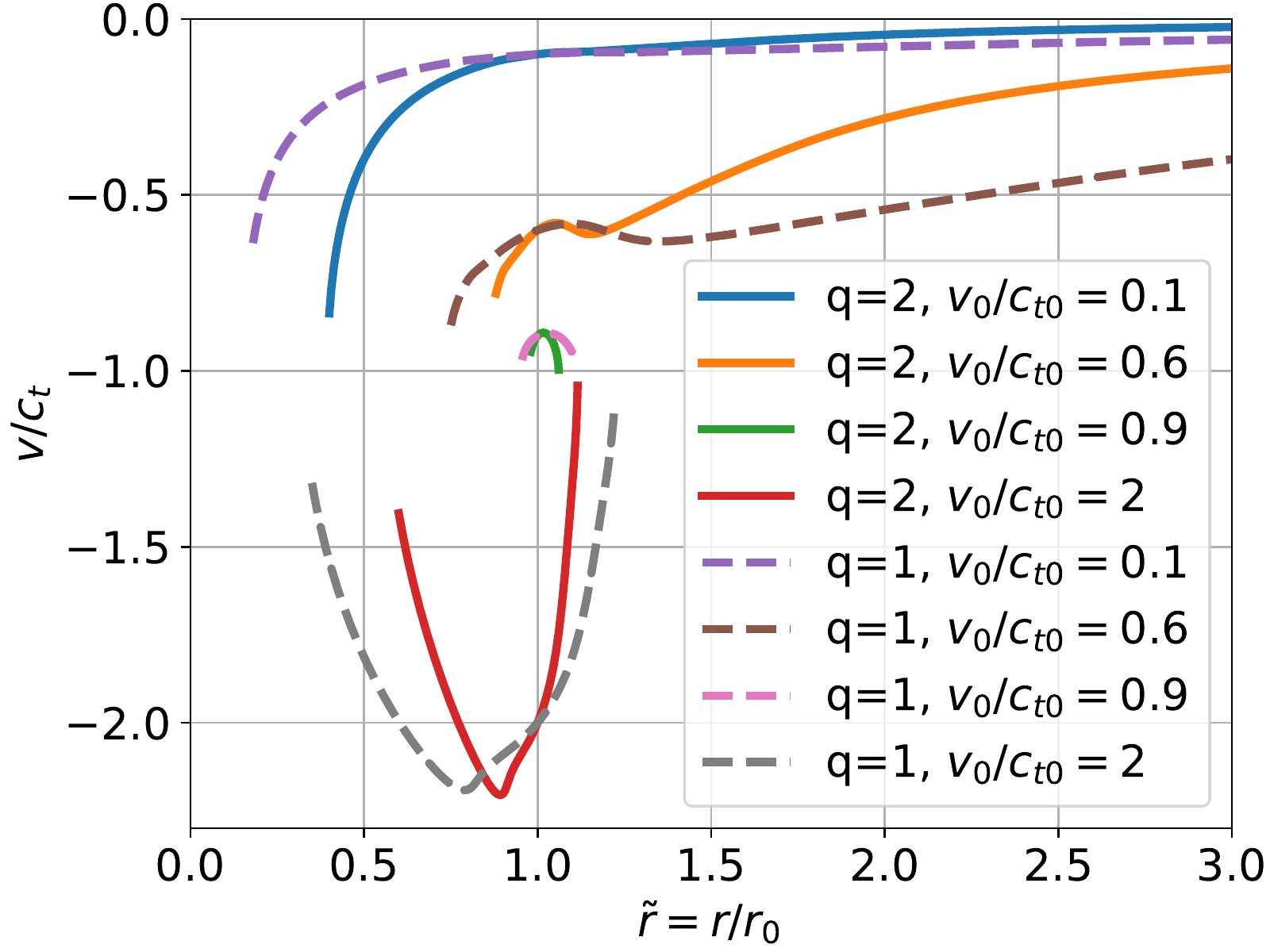}
	\caption [Mach vs r MHD]{Mach number as a function of the scaled radius for MHD steady cooling flow solution for spherical (q=2) and cylindrical (q=1) geometries. We choose $\rm \gamma_m=4/3$, and the parameters at the critical point are $\rm T_0=4\times 10^4$ K, $\rm n_0 =10^{-4}$ cm$^{-3}$, and $\beta_0=0.5$. Note that a transonic solution is not possible in this case, unlike in pure hydrodynamics (see Figure~\ref{fig:discriminant_MHD}). The range of valid steady solutions also decreases closer to the transonic condition.}
	\label{fig:Machvsr_MHD}
\end{figure}

The dimensionless equations with magnetic fields, in a vector form, are
\be
\label{eq:matrix_eqs_mag}
\rm \frac{d}{d\tilde{r}} \begin{bmatrix} \rm \tilde{v} \\ \rm \tilde{s} \end{bmatrix} = \begin{bmatrix} \rm q \left(\frac{c_{s0}}{v_0}\right)^2 \frac{ \left[\frac{\tilde{\rho} \tilde{\Lambda}}{\tilde{v}} \left(1+\frac{\gamma_m}{\gamma \beta_0}\right) + \frac{\tilde{p}_{\rm gas}}{\tilde{r}\tilde{\rho}} \left(1+\frac{\gamma_m \tilde{\rho}^{\gamma_m}}{\gamma \beta_0 \tilde{p}_{\rm gas}}\right) \right] } {
\left[ 1 - \left(\frac{c_{s0}}{v_0}\right)^2\frac{\tilde{p}_{\rm gas}}{\tilde{\rho}\tilde{v}^2} \left(1+\frac{\gamma_m \tilde{\rho}^{\gamma_m}}{\gamma \beta_0 \tilde{p}_{\rm gas}}\right) \right] \tilde{v} } \\ \rm -q \gamma \left(1+\frac{\gamma _m}{\gamma \beta _0} \right) \tilde{\Lambda} \tilde{\rho}^{(2-\gamma)}/\tilde{v}  \end{bmatrix}.
\ee
The entropy equation remains unchanged except for the factor $\rm \left(1+\frac{\gamma _m}{\gamma \beta _0} \right)$ which comes from the scaling of distance by the critical radius. We solve this system of equations similar to the pure hydrodynamics case illustrated before.\footnote{In Appendix \ref{sec:Appendix_A}, we discuss the nature of the solution near the sonic point for transonic solutions using L\'{}H\^{o}pital's rule as before. There we also derive the condition on $\Lambda_T$ for the existence of a transonic solution for a given $\beta_0$.}

Figure \ref{fig:Machvsr_MHD} shows some representative MHD cooling flow solutions with the plasma-$\rm \beta$ at the sonic point $\rm \beta_0=0.5$ and $\rm \gamma_m=4/3$; all other parameters are as in the hydro solutions shown in Figure \ref{fig:Machvsr}. Notice that there is no transonic solution for these parameters, as indicated by the missing $\rm v_0 / c_{t0} = 1$ case. Furthermore, the range of allowed solutions shrinks as we approach the transonic condition. In particular, the $\rm \mathcal{M}_0 = 0.9$ cases (green and pink lines) have no stable solutions outside of a small region surrounding $\tilde{r}=1$.

\section{Numerical Verification with PLUTO}
\label{sec:num_sims}

\begin{figure*}
    \centering
\includegraphics[width=\textwidth]{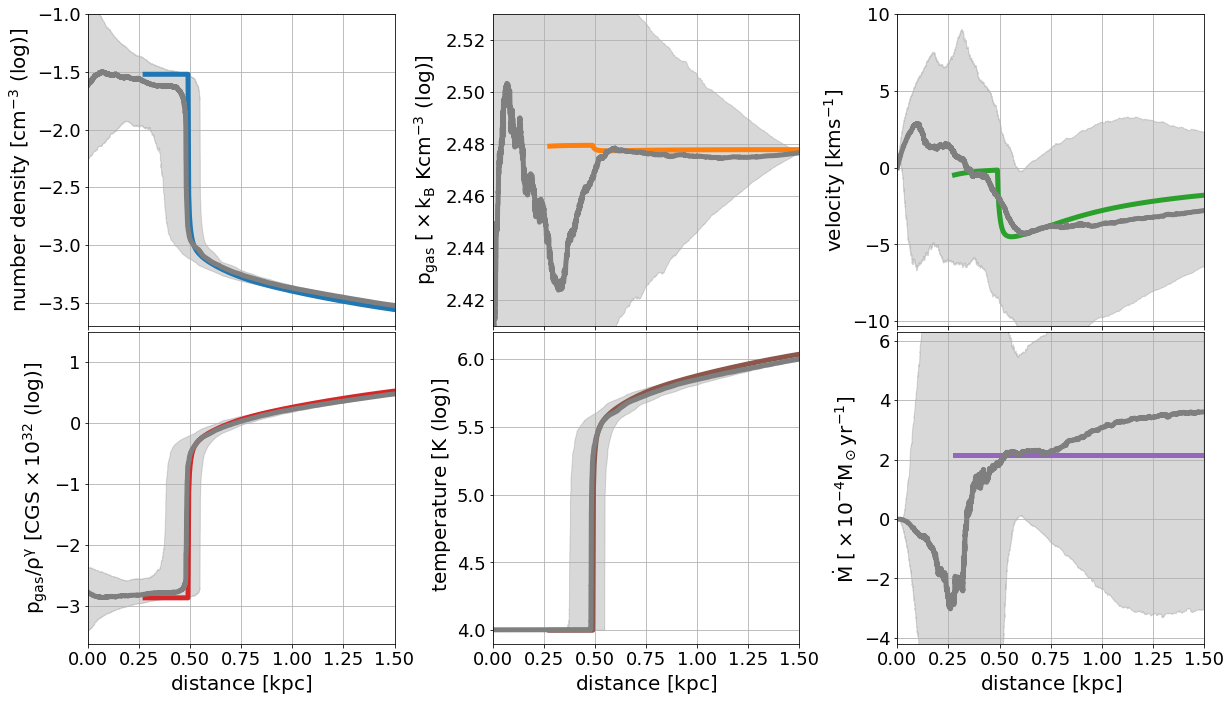}
\caption{The gray curves show the median profiles of number density, pressure, velocity, entropy, temperature, and mass inflow rate over all the time snapshots between $\rm 0.7\ t_{cool,hot}$ and $\rm 3\ t_{cool,hot}$ ($\rm t_{cool,hot}$ is the cooling time for the fixed density and temperature at the outer boundary, $\rm 3\times 10^{-4}\ cm^{-3}$ and $\rm 10^6\ K$ respectively).  
The gray shaded regions indicate the spread between $\rm 16$ and $\rm 84$ percentile values of the respective quantities. 
Our initial/boundary conditions naturally develop a steady, subsonic local cooling flow. The colored curves indicate 
the steady cooling flow solution obtained by solving the ODEs (Eqs. \ref{eq:matrix_eqs}). The 
values used at the critical point to obtain this particular solution of the spherical ($\rm q=2$) cooling flow ODEs are $\rm T_0 = 3.8\times 10^{5} K$, $\rm n_0=7.9\times 10^{-4} cm^{-3}$ and a Mach number $\rm \mathcal{M}_0 = 0.049$, giving a critical radius of approximately $ 560\  \rm pc$. 
The time dependent profiles have fluctuations due to acoustic oscillations, the strength of which is determined by the ratio $\rm t_{cool}/t_{sc}$ where $\rm t_{sc}$ is the sound-crossing time across the entire domain and $\rm t_{cool}$ is the gas cooling time. Since the pressure is close to isobaric and velocity is subsonic, these fluctuations are most visible in pressure and velocity.
}
\label{fig:sphr-pde_compare}
\end{figure*}

In this section, we test our cooling flow ODE solutions with a hydrodynamic PDE solver that evolves the 1D Euler equations with radiative cooling. We have considered the solution in both spherical and cylindrical coordinates, but here we only show a comparison with the spherical solution; the level of agreement in a cylindrical geometry is similar. 

The PDE hydro solutions used in this work are obtained using the PLUTO code (version $\rm 4.4$; \citealt{2007ApJS..170..228M}) which implements a finite volume Godunov-type Riemann solver to numerically solve the equations of magnetohydrodynamics in their conservative form (see Table \ref{tab:solver-config} for details). We do not include any magnetic fields due to the ambiguity in setting up the initial magnetic field configuration. Recall that the MHD equations 
evolve the magnetic field vector rather than a polytropic magnetic pressure for which a somewhat adhoc $\rm \gamma _m$ needs to be chosen.

We initialize the gas profiles in spherical geometry with uniform density and temperature . Radiative cooling is turned off below a gas temperature of $\rm 10 ^4 K$ in both our ODE solver and the PLUTO hydro-solver. The outer density and temperature are kept fixed to $10^{-3.5}$ cm$^{-3}$ and $10^6$ K, respectively, corresponding to the hot CGM. The velocities in the outer ghost zones are copied from the last active zone. The boundary condition is set to inflow-outflow at the inner boundary. Fixing the outermost density and temperature mimics the ambient hot gas CGM in rough thermal and hydrostatic equilibrium. The values chosen for the boundary temperature and density are typical for the hot, volume-filling component of the CGM. Analysis of the temperature and density distribution of the CGM gas (see Appendix ~\ref{sec:Appendix_C}) shows the presence of this hot/intermediate phase across a wide range of redshifts, for a time exceeding the cooling time of this phase ($\sim 1$ Gyr). This gas reservoir is maintained by other effects that we do not model, e.g., feedback, and external gravity. Fixing the outer density/temperature allows us to account for the hot ambient CGM in our local modeling.   
\begin{table}
	\centering
	\caption{The configuration of our PLUTO simulations.}
	\label{tab:numerical-solver}
	\begin{tabular}{lccr}
		\hline
		\hline
		Geometry & Spherical \\ 
		Solver & HLLC \\
		Cooling & Townsend (Solar metallicity)\\
		Code units & $\rm kpc,\ km\ s^{-1}, m_p\ cm^{-3}$\\
		Domain (code units) & 0.01 to 1500 \\
		Spatial Resolution & 128 to 32768 equal volume grid cells \\
		Reconstruction & Parabolic\\
		Time stepping & RK3\\
		Equation of state & Ideal gas\\
		CFL value & 0.3\\
		\hline
	\end{tabular}
	\label{tab:solver-config}
\end{table}

The results of this test are illustrated in Figure~\ref{fig:sphr-pde_compare}. After a cooling time (which is uniform initially), the inner gas cools and becomes denser, flowing in. Since the outer radius has a fixed temperature/density and the inner pressure is smaller than the outer one due to cooling, a pressure-driven steady cooling flow is established after a few cooling times. On top of the steady average profiles (shown by solid gray lines), there are persistent acoustic fluctuations (indicated by the shaded gray regions). We have verified that these fluctuations are much smaller if we keep the inner pressure and density fixed to the steady solution. These fluctuations are signatures of acoustic pulsations in over-dense clouds about a quasi-steady cooling flow; they are persistent, and may be similar to those observed by \citealt{2020MNRAS.494L..27G}. 
Our time-averaged profiles compare favorably with the steady ODE cooling flow solutions. The PLUTO-generated profiles (in gray) in Figure~\ref{fig:sphr-pde_compare} show our highest resolution run, compared to other simulations with identical initial and boundary conditions that we used for our convergence study (Appendix~\ref{sec:Appendix_B}). 


\section{Comparison with Cold Clouds in the TNG50 Cosmological Simulation}
\label{sec:tng}

Here we compare our cooling flow model against the properties of cold clouds within halos in a cosmological MHD simulation. In particular, we assess the structure and cooling properties of small ($\sim$kpc) cold clouds found to populate high-mass halos by the thousands \citep{Nelson2020}, similar to the inferred large abundance of cold gas surrounding luminous red galaxies (LRGs) in SDSS \citep{Anand2021}. \citet{Nelson2020} concluded that the cool phase of the CGM results from cooling on to the dense, cold `seeds' of gas primarily produced due to the strong density perturbations of the satellite galaxies.

These halos are formed within the TNG50 simulation\footnote{\url{https://www.tng-project.org}} (\citealt{pillepich19,nelson19b}) which is the highest resolution run of the IllustrisTNG galaxy formation model \citep{weinberger17,pillepich18}. It simulates a representative $\sim 50$ Mpc comoving side-length volume of the Universe with a baryonic mass resolution of $\sim 8\times10^4 \rm{M}_\odot$, a median spatial resolution of $\sim 150$ parsecs in the star-forming ISM, decreasing to better than 2 kpc within the virial radius of such massive ($\gtrsim 10^{13} M_\odot$) halos. TNG50 has shown diverse manifestations of hydrodynamical phenomenon related to gaseous halos, including the aforementioned cold phase clouds, the production of Lyman-alpha halos at high-redshift \citep{Byrohl2021}, the generation of outflow-driven bubbles around Milky Way and M31-like galaxies similar to the Fermi bubbles \citep{Pillepich2021}, ultraviolet metal-line emission from MgII in the CGM \citep{Nelson2021}, and observable predictions for an azimuthal angle modulation of CGM metallicity \citep{Peroux2020} as well as satellite galaxy quenching \citep{MartinNavarro2021}.

\begin{figure}
    \centering
    \includegraphics[width=\columnwidth]{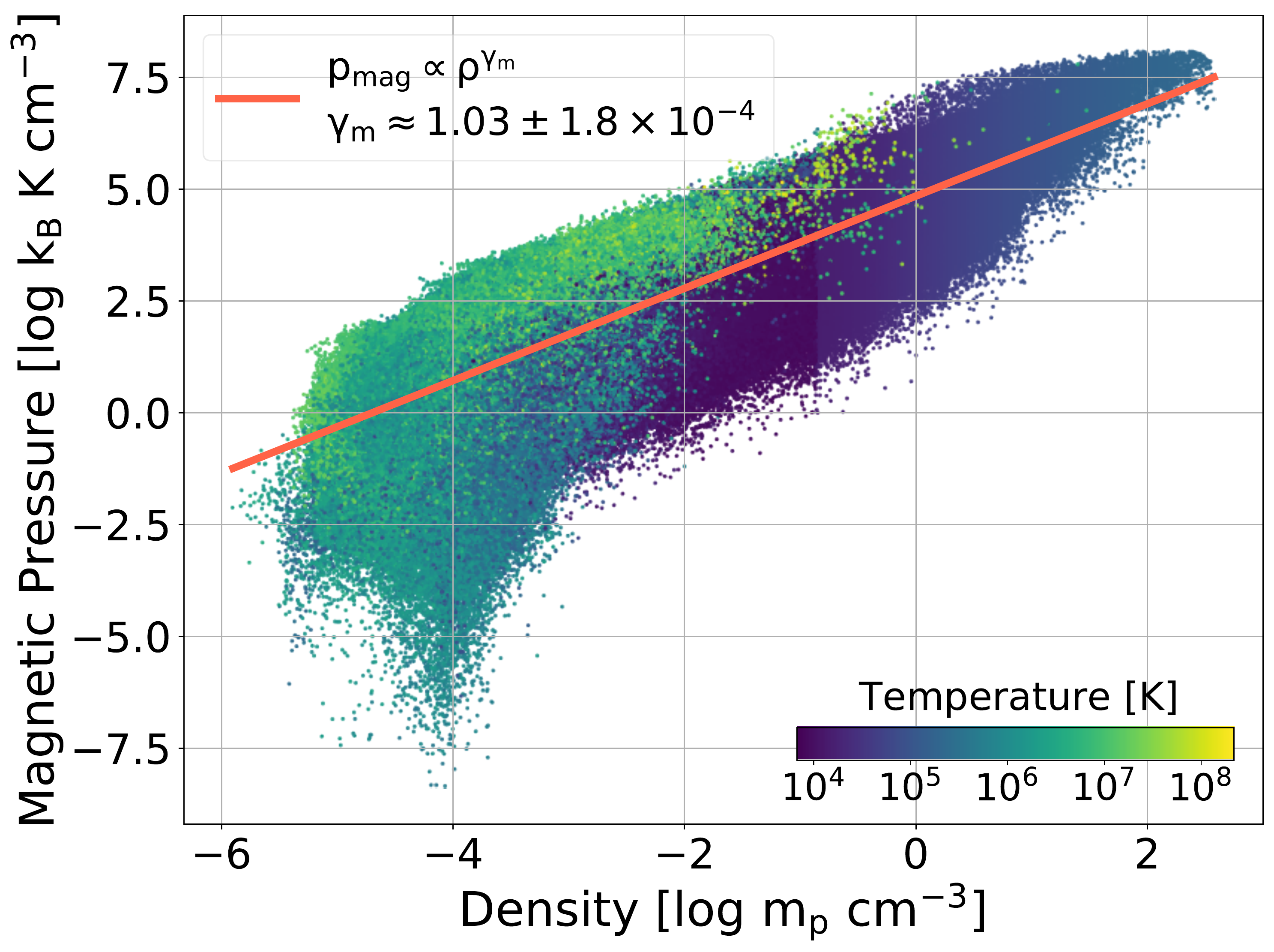}
    \caption{Scatter plot of the magnetic pressure versus gas density for all the cells within our fiducial massive halo (ID 8 at $z=0.5$) in TNG50. The color of points indicates gas temperature, and the red solid line shows a power-law fit using all data points. Although this magnetic pressure-density relation has a significant scatter, the best-fit adiabatic index of $\rm \gamma _m \approx 1.03$ captures the     overall behavior well. Excluding gas cells with non-zero star formation rate removes those with density $> 0.1$ cm$^{-3}$, but increases $\rm \gamma_m$ only by 2\%. Thus, we choose $\rm \gamma_m = 1.03$ for our MHD cooling flow solutions when comparing with TNG50 clouds.} 
    \label{fig:mag-indx}
\end{figure}

\begin{figure*}
    \centering
    \includegraphics[width=\textwidth]{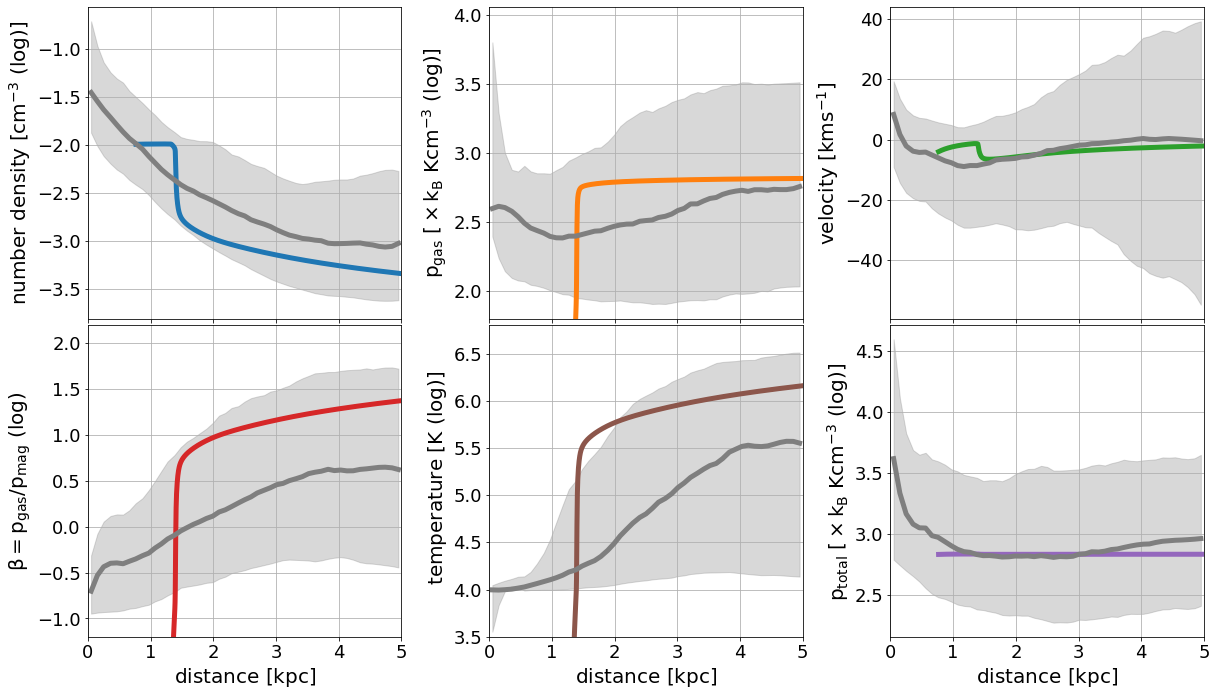}
    \caption{Comparison of analytical cooling flow solutions, including magnetic fields, with stacked cloud profiles extracted from the TNG50 cosmological simulation. Here we include clouds with MgII number density $>10^{-8}$ cm$^{-3}$ and sizes between $\rm 1.0$ and $\rm 1.5$ $\rm kpc$ (303 clouds) from a single massive halo ($\sim 10^{13} M_\odot$; ID 8, $z=0.5$). Gray lines show the median profiles and the shaded portions the $1\sigma$ spread. The gas profiles from our solution (colored lines, $\rm q=2$ spherical geometry) are contrasted against the stacked profiles for TNG50 clouds. To make this comparison we made a by-eye selection of the values for our model free parameters, such that the outcome was reasonably consistent with the TNG50 cloud profiles. The temperature, number density, Mach number and magnetic to gas pressure ratio $\rm \beta$ at the critical point $r_0$ (=1.35 kpc, determined by Eq. \ref{eq:r0_MHD}) are taken to be $\rm 2.0 \times 10^3\ K$, $\rm 1.0 \times 10 ^{-2} \ cm^{-3}$, $\rm 0.2$ and $\rm 0.03$, respectively. The profile behaviors are qualitatively similar, especially the accelerating inflow that slows down near the cloud boundary (green), and the weak pressure gradient that drives the subsonic cooling flow (purple). 
    }
    \label{fig:tngcomp}
\end{figure*}

In TNG50 the large populations of cold clouds form and exist within the hot atmospheres of large halos. Figure 9 of \citealt{Nelson2020} shows the distribution of cloud properties -- namely, radius, metallicity, halo-centric velocity, and halocentric distance -- within a single massive ($10^{13.9} M_\odot$) halo. Similarly, their Figure 10 shows the internal structural properties of cold clouds, by analyzing median radial profiles obtained by stacking together clouds in different radius bins for a $\sim 10^{13} M_\odot$ halo.\footnote{TNG50-1 \href{https://www.tng-project.org/api/TNG50-1/snapshots/67/halos/8/}{halo ID 8} at snapshot 67; see the \href{https://www.tng-project.org/api/TNG50-1/snapshots/67/halos/8/vis.png?partType=gas&partField=density&size=5.2&sizeType=rVirial&method=sphMap&nPixels=2048\%2C2048&axesUnits=kpc&relCoords=True&plotStyle=edged&labelZ=True&labelScale=True&labelSim=True&labelHalo=True&title=True&colorbars=True&ctName=viridis}{gas density image}.} Using the catalog of identified clouds for this halo, we extract similar \textit{median} profiles of cold clouds, and compare with our steady-state cooling flow solutions. Note that TNG50 includes magnetic fields, which were found to dominate the total pressure with $\beta \ll 1$ inside cold clouds. In all our subsequent analysis of the TNG50 data, unless otherwise explicitly stated, we have deliberately excluded a small number of gas cells that have non-zero star formation rates. This allows us to eliminate the ISM pressure contribution and focus on the CGM gas.

Figure~\ref{fig:mag-indx} shows a scatter plot of magnetic pressure versus gas density for individual gas cells in the simulation. The color of the data points shows the gas temperature, and we see that the highest magnetic pressure occurs in the coolest/densest cells. In addition, a polytropic equation of state for magnetic pressure (see Eq. \ref{eq:p_mag}; red line) is a reasonable approximation. To capture the impact of magnetic fields in the cooling flow solutions we therefore adopt a best-fit polytropic index for magnetic pressure of $\rm \gamma_m=1.03(\pm 1.8 \times 10 ^{-4})$. Note that the quoted error is statistical and lower than the total scatter.

We need to choose the parameters of our cooling flow model to compare with clouds in TNG50. These are the density, temperature, and Mach number at the subsonic critical point. To do so, we select by-eye values roughly consistent with the median profiles around TNG50 clouds, omitting any systematic search for best-fit parameters. The result is shown in Figure~\ref{fig:tngcomp} in terms of six radial profiles: number density, gas pressure, velocity, plasma-$\rm \beta$, gas temperature, and total pressure. We compare our cooling flow solutions (colored lines, $\rm q=2$ for spherical geometry) with the stacked, median TNG50 cloud profiles (gray lines, and $\rm 1\sigma$ cloud to cloud variation as the shaded band). The profiles around individual clouds have a large scatter (as indicated by the shaded band) which is averaged out on stacking.

Here we focus on clouds with radii between $1.0$ and $1.5$ kpc (unless mentioned otherwise, all our distances correspond to physical rather than comoving units), but note that similar results hold for other cloud sizes. Our parameter choice for the presented solutions is $\rm T_0=2\times 10^3$ K, $\rm n_0=0.01$ cm$^{-3}$, $\rm v_0/ \rm c_{t0}=0.2$ and $\rm \beta_0=0.03$, which also fixes the critical radius ($\rm r_0=1.35$ kpc for our parameters; see Eq. \ref{eq:r0_MHD}). This particular solution corresponds to a steady cold gas mass inflow rate of $\approx \rm 5.2 \times 10^{-4} M_\odot \ yr^{-1}$.

Overall, we find that the analytical solutions roughly follow the cloud profiles seen in TNG50. Our solutions generally fall within the spread of stacked profiles, although there is substantial deviation from the medians. The largest discrepancies are found within clouds themselves, which is expected as we do not model any gas dynamical effects at the centers of clouds. Although a systematic search for best-fit model profiles may improve the level of agreement, the analytic profiles are not expected to quantitatively match the simulations, as we discuss next.

\subsection{Limitations and applicability of the models}

There are several physical reasons why our analytical profiles and the stacked profiles around TNG50 clouds differ. First, the cooling function in TNG50 depends not only on the temperature (see Eq. \ref{eq:ent}) but also the local metallicity, density, and the UV background coupled with a local AGN radiation source, both subject to self-shielding. Second, the stacking of multiple clouds of different sizes leads to smearing of sharp features in the median. Third, the cold clouds in TNG50 are not spherically symmetric because they are not at rest, but instead are moving through the hot CGM of their host halo. Finally, the numerical resolution of TNG50 is necessarily finite, and the gas dynamics at the smallest/cloud scales will not be resolved. The critical radius ($\rm r_0$; see Eq. \ref{eq:r0_MHD}), an important length scale in our model, is often smaller than the available numerical resolution, especially for the densest clouds.

Nevertheless, the ability of our solutions to reproduce the qualitative behavior of TNG50 cold clouds suggests that the cooling induced pressure gradient plays an essential role in setting the local environment of CGM clouds. Localized turbulence near the cloud-CGM interface can potentially lead to deviations of the TNG50 results from our steady cooling flow model. There can be significant turbulent transport of mass/momentum/energy between clouds and their surroundings \citep{Fielding2020}, whereby the cooling flow description breaks down. Such turbulent transport can be modeled via a mixing-length prescription (\citealt{Tan2021}), which is however beyond the scope of this work.

\begin{figure}
    \centering
    \includegraphics[width=0.98\columnwidth]{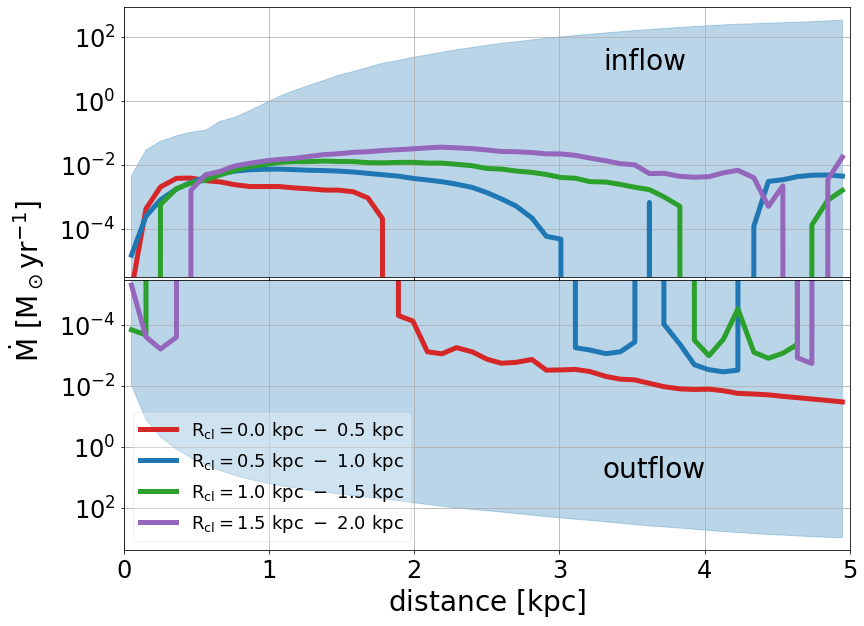}
    \caption{Magnitude of gas mass inflow and outflow rates for cold clouds in four different size bins from our fiducial massive halo of TNG50. The median mass flow rate ($\dot{\rm M} \rm = -4 \pi r^2 \rho v$, for spherically symmetric flow) is calculated using the median values of fluid variables from our stacked cloud profiles (924, 764, 303, and 133 total clouds used in the four bins, respectively). The shaded area denotes the $1\sigma$ scatter for the cloud size (1-1.5 kpc) we analyze in Figure~\ref{fig:tngcomp}. The other cloud sizes also show a comparable spread but we do not show it for clarity. The largest accretion rate at some radii can be orders of magnitude larger than the median. Both inflowing and outflowing gas show orders of magnitude variation in the flow rates. The mass flow rate, as inferred from median profiles, is not constant with distance, indicating a non-steady flow and/or a substantial turbulent mass transport. 
    }
    \label{fig:Mdot}
\end{figure}

To better understand the dynamical flows on to the clouds, Figure~\ref{fig:Mdot} shows the mass flow rate \mbox{$\dot{\rm M} = - 4 \pi r^2 \rho v$}, derived using the median stacked profiles of TNG50 clouds as a function of radius, with clouds collected in bins based on their size. We caution that the mass accretion rate based on median profiles does not account for turbulent mass flux ($\rm - 4 \pi r^2 \langle \delta \rho \delta v \rangle$) which may be substantial. We see that the mass accretion rate as a function of radius from the cloud is not constant, an assumption made in our steady cooling flow model. This limits the quantitative comparison of the cooling flow solution and the TNG50 cloud profiles. Larger clouds tend to have higher inflow rates. In fact, mass flows outwards from the smallest clouds outside $\sim 2$ kpc, possibly indicating their destruction by turbulent (and/or numerical) mixing/heating. In contrast, our cooling flow model only accounts for cloud growth due to mass inflow.

Figure~\ref{fig:Mdot} shows that both local inflows and outflows with an unsteady nature exist. This produces the wide spread in the mass flow rates when stacking, and may explain why the mass inflow rate inferred from differential emission (discussed extensively in section~\ref{sec:discussion}) is much larger than the one obtained by fitting the radial profiles.


\section{Discussion \& Astrophysical Implications}
\label{sec:discussion}

Our cooling flow model is local and represents the cooling-induced flow around dense clouds in the CGM. Unlike classical cooling flows, there is no external gravity. Dense gas close to clouds cools and drives a pressure gradient towards the center. As a result, there is a net inflow of cooling gas. The diffuse/hot CGM has a long cooling time, while its cooling losses are also compensated in some part by feedback heating, and thus a pressure gradient is maintained. While a detailed study of the impact of feedback is beyond the scope of this work, we show in Appendix~\ref{sec:Appendix_C} that the hot/intermediate temperature gas in the CGM (as measured in TNG50 halos), even with a cooling time shorter than its age, is long-lived. As a result, diffuse hot gas in the CGM is long-lived, compared to the time needed for local cooling flows to develop. However, cooler/denser gas closer to clouds has a short enough cooling time to set up a pressure-driven local cooling flow.

The presence of this long-lived hot reservoir complicates the idea that two gas phases (hot and cold), in steady state, must achieve pressure balance without radiative losses, in absence of any heat transport (say due to thermal conduction; e.g., see \citealt{Tan2021}). Instead, the picture that we present here is that hot gas, with a large volume-filling fraction, is maintained as a mass reservoir due to feedback heating and sustains a cooling flow of denser/cooler gas around dense clouds. In section~\ref{sec:num_sims} we model the impact of this hot reservoir as a fixed density/temperature at the outer boundary.

Multiphase gas is ubiquitous in astrophysical coronae, including the CGM. The key physical ingredients of these clouds embedded in a diffuse medium are radiative cooling, magnetic fields, and boundary-layer turbulence driven by relative motion. Gravity-driven cooling flows have been studied in the context of cool core clusters for decades (for a review see \citealt{Fabian1994}; a recent work is \citealt{Stern2019}; see also  \citealt{Prasad2020} for application to the Phoenix cluster). Here we have studied analogous flows around cold clumps driven only by radiative cooling. The mass cooling rate ($\dot{\rm M}_{\rm cool}$) and the radiative cooling rate ($\dot{\rm E}_{\rm cool}$) are proportional in a classic cooling flow. Here we generalize this relation to include strong magnetic fields, fast flows, and background gravity. 

However, such a close relation between the mass cooling/accretion rate and the internal energy loss rate due to radiative cooling breaks down in the turbulent boundary layers around CGM clouds. In fact, the temperature distribution of gas with radiative losses in a homogeneous cooling flow is fundamentally different from a radiative mixing layer (the fundamental building block of the multiphase CGM); as argued in \citet{Kanjilal2021} in the context of cloud-crushing simulations. 

\subsection{Differential emission from gas in and around cold clouds}\label{sec:dif-emm}

To determine the temperature regimes which exhibit steady cooling flow like behavior, we consider 
a steady one dimensional cooling flow. In our cooling flow model, we include a simple model of magnetic fields by treating it as a polytropic fluid. Therefore, the internal energy equation for the ``magnetic" gas satisfies
$$
\rm \rho \frac{d \epsilon_{mag}}{dt} =  -p_{mag} {\bf{\nabla \cdot v}}, 
$$
where $\rm \epsilon_{mag}= \rm (p_{\rm mag}/\rho)/(\gamma_m-1)$ is the specific internal energy of the magnetic fluid and $\rm d/dt$ denotes the Lagrangian derivative. The total energy equation for the system (including a time-independent gravitational potential $\Phi$, which is trivial to include but is not considered in this work) is
\be
\rm \partial_t \left[\rho \left( \epsilon + \frac{1}{2} v^2 + \Phi \right)\right] + {\bf{\nabla}.}\left[\rho \left( \epsilon +  \frac{p}{\rho} + \frac{1}{2} v^2 + \Phi \right) {\bf{v}}\right] \\
\rm  = 
- n_e n_i \Lambda(T),
\ee
where $\rm \epsilon = \epsilon_{\rm gas} + \epsilon_{\rm mag}$, $\rm p = p_{\rm gas} + p_{\rm mag}$, $\rm p_{\rm gas} = \epsilon_{\rm gas}/[\rho (\gamma-1)]$ and $ \rm p_{\rm mag} = \epsilon_{\rm mag}/[\rho (\gamma_m-1)]$.

In a steady ($\rm \partial_t=0$) one-dimensional flow, the above equation reduces to
$$
\rm \frac{1}{r^q}\frac{d}{dr}\left[\rho v r^q \left( \frac{\gamma}{\gamma -1}\frac{p_{gas}}{\rho} + \frac{\gamma _m}{\gamma _m -1} \frac{p_{mag}}{\rho}+\frac{1}{2} v^2 + \Phi\right)\right]= - n_e n_i \Lambda (T),
$$
which applies even for the full MHD equations, in addition to the polytropic assumption we have made in this work. Radiative cooling implies that
$$ 
\frac{{\rm d} \dot{\rm E}_{\rm cool} }{\rm dr}=\rm n_e n_i \Lambda (T) K r^q,
$$ 
and adopting $\dot {\rm M} = \rm - K r^q \rho v$ for the mass influx rate due to cooling, one obtains the following expression for the radiative luminosity as a function of gas temperature (i.e. the differential emission),
\begin{equation} 
\label{eq:SCF_diffemm}
\frac{{\rm d}\dot{\rm E}_{\rm cool}}{\rm d \log_{10} T} = \dot{{\rm M}}_{\rm cool} \rm \frac{d {\cal B}}{d \log_{10} T}.
\end{equation}
Here we assume a one-to-one relation between radius and gas temperature, as well as a constant mass cooling rate, which equals the mass accretion rate, on to a cold cloud. The Bernoulli number $\rm {\cal B}$ is given by 
\be
\label{eq:Bernoulli}
\rm {\cal B} = \left(\frac{\gamma}{\gamma -1} + \frac{\gamma_m}{\gamma_m -1} \frac{1}{\beta} \right) \frac{k_B T}{\mu m_p} +\frac{1}{2} v^2 + \Phi.
\ee
This is a generalization of the classic cooling flow model in the context of galaxy clusters (\citealt{Fabian1994}) for which ${\rm d}{\dot{\rm E}_{\rm cool}}/{\rm d T} = (5/2)\dot{\rm M}_{\rm cool} \rm k_B/(\mu m_p)$ and applies exactly for a homogeneous \mbox{1-D} cooling flow, irrespective of the flow geometry. As a sanity check we have verified that the relation given by Eq. \ref{eq:SCF_diffemm} holds exactly in our cooling flow solutions, in both the hydrodynamical and magnetic field cases.


For pure hydrodynamical steady cooling of uniform gas around clouds, in the absence of gravity, ($\rm v \ll c_s $, $\rm \beta \rightarrow \infty$ and $\Phi = 0$), the Bernoulli number reduces to $\rm {\cal B} = \gamma k_B T/[(\gamma-1)\mu m_p]$ and one obtains the standard expression for differential emission \citep{Kanjilal2021},\footnote{We note that $ {\rm d} \dot{\rm M}_{\rm cool}/{\rm d \log_{10} \rm T}$ in Eq. 12 of \citealt{Kanjilal2021} is incorrect and should actually be $\dot{\rm M}_{\rm cool}$.} 
\begin{equation} 
\label{eq:CF_diffemm}
\frac{{\rm d}\dot{\rm E}_{\rm cool}}{\rm d \log_{10} T} {\rm \approx 5 \times 10^{33} {\rm~erg~ s}^{-1} \left( \frac{T}{10^4{\rm~K}} \right)} \left( \frac{\dot{\rm M}_{\rm cool}\rm }{10^{-5} \rm M_\odot {\rm yr}^{-1}} \right).
\end{equation}

\begin{figure}
    \centering
    \includegraphics[width=\columnwidth]{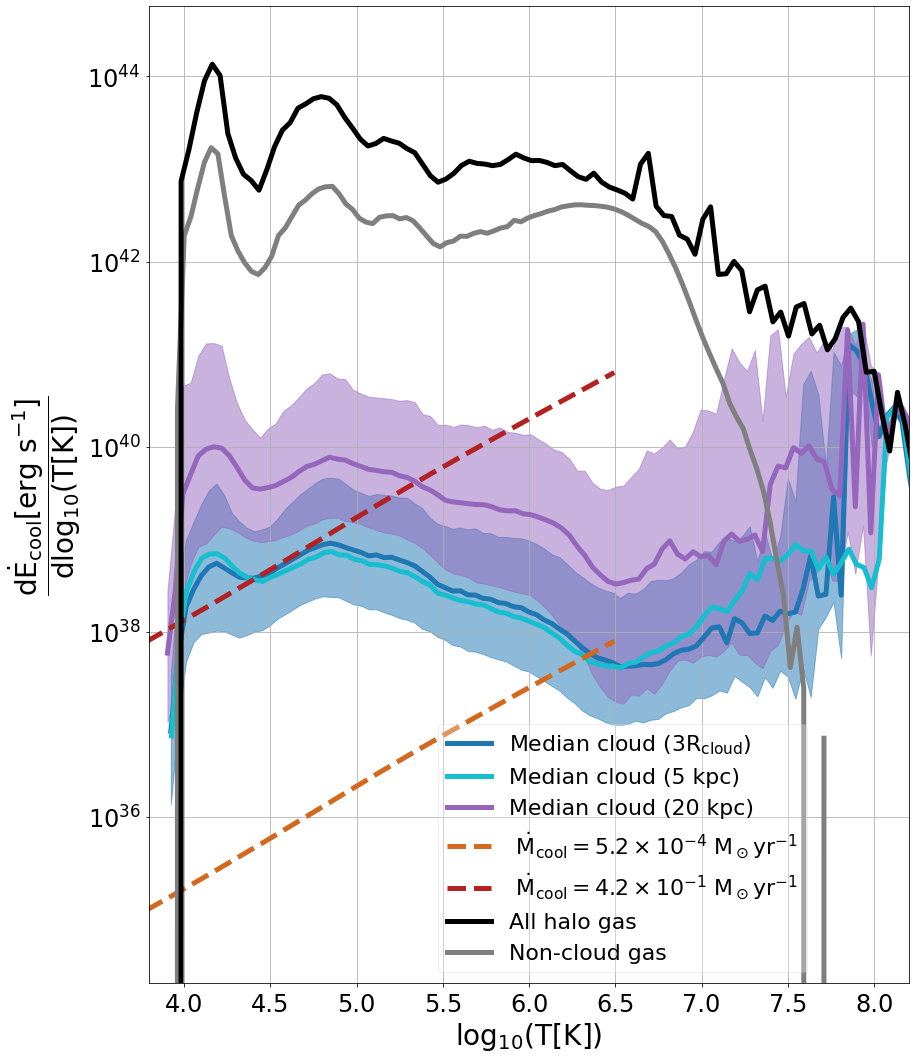}
    \caption{Differential emission as a function of temperature. The solid colored lines show median value from all gas within $\rm 3$ times the respective cloud size (blue), $\rm 5\ kpc$ (cyan) and $\rm 20\ kpc$ (purple) from the centers of cold clouds within our fiducial TNG50 halo. Here we stack together all $\approx 1.4 \times 10^4$ clouds identified (with $n_{\rm MgII}>10^{-8}$ cm$^{-3}$) in the halo. The shaded bands show the corresponding $1\sigma$ scatter. The dashed orange line shows the analytic expression from the steady cooling flow model (using Eq.~\ref{eq:SCF_diffemm} with the parameters from Figure~\ref{fig:tngcomp}), which correspond to a mass cooling rate ($\dot{\rm M}_{\rm cool}=5.2 \times 10^{-4} \rm M_\odot yr^{-1}$). The red dashed line instead corresponds to $\dot{\rm M}_{\rm cool}=4.2 \times 10^{-1} \rm M_\odot yr^{-1}$, a much higher value than obtained from fitting the stacked profiles (see text). The black line shows the emission from the entire halo, while the gray line omits gas that is within $3$ times the cloud radius of any cloud. The gray curve shows a bump at $10^{6.5}$ K corresponding to the volume-filling hot gas at the virial temperature.
    }
    \label{fig:diff-emm}
\end{figure}

Figure~\ref{fig:diff-emm} shows the differential emission as a function of temperature from the gas in and around cold clouds from our fiducial TNG50 halo. Both median differential emission profiles within 5 kpc and $3~{\rm R_{\rm cloud}}$ give a similar result. We note that, on average, there is one neighboring cloud within this radius around each cloud, such that the median emission should be an overestimate by $\sim 2$. We find that the differential emission due to a steady cooling flow is only approximately valid within a limited range $\sim 0.5 \,\rm{dex}$ in temperature around $\sim 10^{4.5}$ K. The luminosity as a function of temperature is qualitatively similar to that from radiative cloud-crushing simulations (see Figure 6 in \citealt{Kanjilal2021}), in that the emission falls slowly as a function of temperature toward intermediate temperatures rather than rising linearly, as predicted by a steady cooling flow model. 

We note that the cooling flow profiles in Figure~\ref{fig:tngcomp} gives a mass inflow rate of $\dot{\rm M}_{\rm cool}=5.2 \times 10^{-4} \rm M_\odot yr^{-1}$ (a comparison of median and cooling flow profiles in Figure~\ref{fig:tngcomp} shows that this is an underestimate by a factor of few). However, we find that a much higher mass inflow rate $\dot{\rm M}_{\rm cool} \approx 4.2 \times 10^{-1} \rm M_\odot yr^{-1}$ better matches the mass cooling/inflow rate inferred from Figure \ref{fig:diff-emm} (using Eq. \ref{eq:SCF_diffemm}) at the temperature phases where radiative cooling is most efficient (Figure 7 in \citealt{Kanjilal2021} shows that the isobaric cooling time is the shortest at $\sim 2 \times 10^4$ K). 

\begin{figure}
    \centering
    \includegraphics[width=\columnwidth]{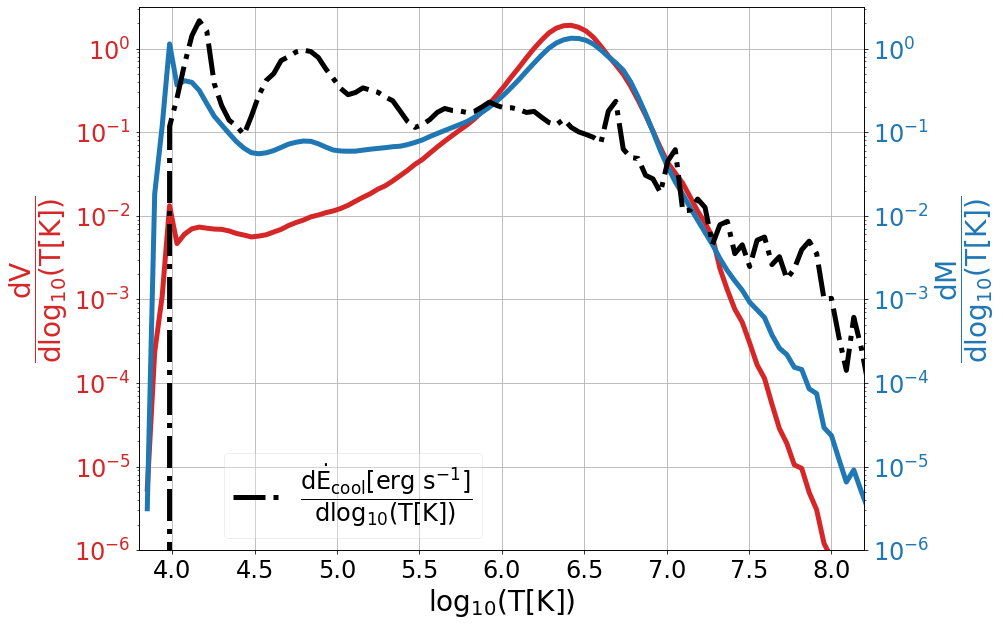}
    \caption{The mass (blue solid line), volume (red solid line), and emissivity (black dot-dashed line) weighted normalized distribution of all the halo gas. The virial temperature of the halo is quite prominent in the mass and the volume weighted PDFs. The emissivity PDF is dominated by dense clouds and the virial temperature bump is almost fully washed out. On the other hand, as illustrated in Figure~\ref{fig:diff-emm}, emission from the volume-filling gas outside the clouds traces the virial temperature of the halo.}
    \label{fig:diff-emm-compare}
\end{figure}

We can understand this discrepancy as follows. The stacked profiles around clouds are volume-weighted averages, but the emission is dominated by denser/cooler gas, which leads to a much larger mass cooling rate based on differential emission as compared to that derived from the cooling flow fit to stacked profiles. Moreover, the profiles around individual clouds have a large scatter as illustrated in Figure~\ref{fig:Mdot}, and this gives rise to a large difference between emission and volume weighted estimates. To illustrate this difference, Figure \ref{fig:diff-emm-compare} shows the normalized temperature PDF of our fiducial TNG50 halo weighted by volume (red line), mass (blue line) and emissivity (dot-dashed line). Note that the cooler temperatures dominate increasingly as we go from volume to mass to emissivity weighted PDFs. In fact, the peak at $\sim 10^{6.5}$ K, corresponding to the halo virial temperature, is entirely smoothed out in the emissivity PDF. Also note that the emission-weighted PDF at $\sim 10^4$ K is more than 100 times larger than the volume weighted PDF, explaining the much higher $\dot{\rm M}_{\rm cool}$ obtained from the emission measure as compared to the stacked profiles. This highlights the difficulties in obtaining physical parameters from a multiphase CGM. 

In a steady cooling flow, cooling rate and mass inflow rate are strictly proportional. However, in TNG50 clouds (and indeed in nature), turbulence may dominate the movement of gas across temperatures, especially in phases with relatively inefficient radiative cooling at increasingly higher temperatures. At these temperatures the linear relation between $\dot{\rm M}_{\rm cool}$ and $\dot{\rm E}_{\rm cool}$ breaks down. This behavior of luminosity as a function of temperatures seems generic to all radiative multiphase turbulent flows, from multiphase cool core clusters (Figure 3 in \citealt{Sharma2012}) to radiative layers around cool filaments in hot halos (Figure 14 in \citealt{Mandelker2020}). This similarity suggests that radiative mixing layers are a fundamental building block of the complex multiphase CGM, and that it is not well described by a steady single-phase cooling flow.

Figure~\ref{fig:diff-emm} also shows the median differential emission from the local environment within $\rm 20\ kpc$ from the center of each cloud. There are around $\rm 1.4 \times 10^4$ clouds in our fiducial halo, and given the volume of the halo ($3.6 \times 10^8$ kpc$^3$), about this many non-overlapping clouds of size $\rm 20\ kpc$ would fill up the entire halo. However, we find that the clouds are highly clustered with typical separations much smaller than $\rm 20\ kpc$. As a result, they are significantly less volume filling, yet play a dominant role in influencing the emission properties of the halo. Even a local environment as small as $\rm 5\ kpc$ around each cloud has significant overlap with multiple clouds.\footnote{Approximately $\rm 3$\% of around $\rm 15$ million non-star forming gas cells in the halo are within just $\rm 5\ kpc$ from the center of one or more clouds. Out of this $\rm 3$\%, about $\rm 11$\% of the cells fall within the $\rm 5\ kpc$ overlapping region of two or more clouds.} Therefore, the median differential emission from within a $\rm 20\ kpc$ local environment around each cloud is much higher as it includes multiple clouds. Clustering of clouds and their slightly lower pressures can lead to their mergers (e.g., see \citealt{Waters2019b,Das2021}), partly compensated by turbulent fragmentation (\citealt{Mohapatra2019}).

Figure \ref{fig:diff-emm} also shows the differential emission from the entire fiducial halo (black line) and from volume elements beyond $\rm 3 R_{\rm cloud}$ of all our clouds (gray line). The emission is dominated by $10^4-10^5$ K gas and not by the virial temperature ($\sim 10^{6.5}$ K) gas. A bump corresponding to the virial temperature gas is visible in the non-cloud gas. This gas also shows significant emission at low temperatures, implying that our cloud selection criterion of $n_{\rm MgII}>10^{-8}$ cm$^{-3}$ misses out a non-negligible volume of cells at $\sim 10^4$ K.

While the CGM of typical star-forming galaxies has often been explored using quasar absorption studies (see \citealt{Tumlinson2017} for a review), emission directly probes the radiative losses and the concomitant flow of mass across temperature phases \citep{bertone12,corlies16,Nelson2021}. However, our work shows that one cannot rely on a simple cooling flow model across all temperature ranges. Such a model only applies locally around dense clouds, and in the narrow temperature range with short cooling times, rather than across the full CGM. For example, \citet{Zhang2021} recently stacked $\rm H\alpha$ emission from the CGM of Milky Way like galaxies in SDSS to estimate the mass cooling rate across $10^4$ K, finding it to be $4-90$ times larger than the star formation rate. They report that most of the $10^4$ K gas from the CGM does not form stars but instead is recycled in the galactic wind. Based on our Figure \ref{fig:diff-emm}, extrapolating the results between the hot ($\sim 10^7$ K) and $10^4$ K phases, we anticipate that the mass flux from $\sim 10^4$ K all the way to star-forming cold molecular phase will similarly vary (a similar interplay of cooling and turbulence occurs in the ISM; e.g., see \citealt{Vazquez2000,Audit2005}). Thus, it may be difficult to directly relate the star formation rate and the cooling rate of gas at $10^4$ K as measured by H$\alpha$ emission.


\section{Conclusions} \label{sec:conclusions}

In this study we present analytic, steady-state solutions for the pressure-driven cooling flows around cold clouds, as may exist in the multiphase circumgalactic medium (CGM) surrounding galaxies. We also compare these solutions with cloud properties in TNG50 cosmological simulation. The key conclusions of this work are:

\begin{enumerate}

\item Steady cooling flows around clouds in spherical and cylindrical geometries allow solutions with a critical point where the velocity satisfies $\rm dv/dr=0$. Transonic solutions are also possible for standard cooling functions at $\rm T \gtrsim 10^4$ K (Figures~\ref{fig:Machvsr} and~\ref{fig:discriminant}).

\item Cold clouds in the CGM are magnetically supported because of flux freezing and compression of cooling gas. Therefore, we incorporate the effects of magnetic fields in our solutions using a polytropic equation of state. A cooling flow with magnetically supported clouds does not admit a transonic solution with the standard cooling function, although subsonic solutions exist. The magnitude of the gas profiles in the subsonic regime are most relevant for the CGM (Figure~\ref{fig:Machvsr_MHD}).

\item We compare our analytical cooling flow model with numerical, one-dimensional, time dependent calculations. We verify the existence and stability of our solutions (Figure~\ref{fig:sphr-pde_compare}) by comparing the time-averaged profiles with the steady-state ODE solutions. We conclude that as long as a large difference exists between the cooling times of the gas near dense clouds, versus the cooling time at global scales, cooling flow features develop in the local gas profiles. This situation develops as a consequence of feedback heating which prevents the cooling of the large scale volume-filling gas (Figure~\ref{fig:dist-evolve}), while gas around locally dense seeds can cool. These cooling flows are robust features, insensitive to the initial conditions, and are noticeable in the time-averaged profiles of the gas (Figures~\ref{fig:sphr-pde_compare} and~\ref{fig:sphr-pde_non-const}). They are distinct from the global cooling flows in a gravitational field commonly discussed in the context of cool core clusters.

\item We compare our cooling flow solution including magnetic fields with the structural profiles and cooling properties of gas in and around cold clouds identified in massive halos within the TNG50 cosmological galaxy formation simulation (Figure~\ref{fig:tngcomp}). While we find solutions which qualitatively reproduce the overall radial profiles of gas density, pressure, temperature, and velocity, there are numerous differences. Most notable are the lack of (spherical) symmetry of TNG50 cold clouds due to their relative motion through the hot halo, and the role of turbulent energy transport around the clouds.

\item We generalize the classic cooling flow solution and relate the differential emission to the mass cooling rate and the Bernoulli number (Eq. \ref{eq:SCF_diffemm}). We show that the classic cooling flow relation between $\dot{\rm M}_{\rm cool}$ and $\dot{\rm E}_{\rm cool}$ does not hold for the boundary layers around the TNG50 cold clouds (Figure~\ref{fig:diff-emm}), except for a very narrow temperature range where the isobaric cooling time is very short ($\sim 10^{4.5}$ K). This violation of the key assumption of a steady cooling flow implies that we cannot generally apply the cooling flow relation between $\dot{\rm M}_{\rm cool}$ and $\dot{\rm E}_{\rm cool}$ to interpret observations of the CGM. 

\item From the analysis of differential emission from the clouds and their surroundings (Figures~\ref{fig:diff-emm} and \ref{fig:diff-emm-compare}) we conclude that the emission properties of halo gas can be dominated by the local environment of cold clouds in the CGM, especially for gas phases between $\rm 10^4\ K$ and $\rm 10^{5}\ K$, which have very efficient radiative cooling. Cooling rather than turbulence is therefore the dominant physics that translates gas across phases at these temperatures. In contrast, turbulence dominates in the radiatively inefficient hotter phases. The hotter volume-filling halo gas contributes less emission, except close to the virial temperature.
\end{enumerate}

Our model highlights some of the key physical effects that generate local gas flows in and around cold clouds. Further work is needed to explore the physics and observational implications of the interplay of cooling and turbulence in the CGM, both in the diffuse phase (e.g., see \citealt{Mohapatra2019}) and in the radiative boundary layers around clouds (e.g., see \citealt{Fielding2020,Tan2021}).

\section{Acknowledgments}

The research of AD is supported by the Prime Minister's Research Fellowship (PMRF) from the Ministry of Education (MoE; formerly MHRD), Govt. of India. AD acknowledges the support from the Max Planck Institute for Astrophysics, Garching for hosting him as a visiting student where this work was initiated. AD also acknowledges the useful discussions with his colleagues Ayan Ghosh and Ritali Ghosh which helped improving the readability of this paper. We acknowledge Prof. Siang Peng Oh at UCSB for useful discussions which greatly improved the work presented in this paper. PS acknowledges a Swarnajayanti fellowship from the Department of Science and Technology (DST/SJF/PSA-03/2016-17), a Humboldt fellowship, and a National Supercomputing Mission (NSM) grant from the Department of Science and Technology, India. DN acknowledges funding from the Deutsche Forschungsgemeinschaft (DFG) through an Emmy Noether Research Group (grant number NE 2441/1-1). Our work has benefited substantially from the computational resources provided by the Max Planck Computing and Data Facility (MPCDF), as well as discussions and talks during the KITP `halo21' program 2020 supported by NSF PHY-1748958.

\section{Data Availability}

We have hosted the codes used in our work 
at a \href{https://github.com/dutta-alankar/cooling-flow-model.git}{Github repository}\footnote{\url{https://github.com/dutta-alankar/cooling-flow-model.git}} for public access. Additional animations and visualizations are available at \url{https://github.com/dutta-alankar/cooling-flow-model/blob/main/animations/}. Any other relevant data associated with this article will be shared on reasonable request to the authors. Additionally, we have hosted a video explaining the work presented in this paper in our \href{https://www.youtube.com/channel/UCPw9R4ReQqAHq8ArgJhFWLA}{IISc Computational Astrophysics} YouTube channel. All the data related to the IllustrisTNG simulations, including TNG50, are publicly available at \url{www.tng-project.org/} \citep{nelson19a}.

\bibliographystyle{mnras}
\bibliography{references}


\appendix

\section{Transonic solution with magnetic fields}
\label{sec:Appendix_A}

\begin{figure}
	\includegraphics[width=\columnwidth]{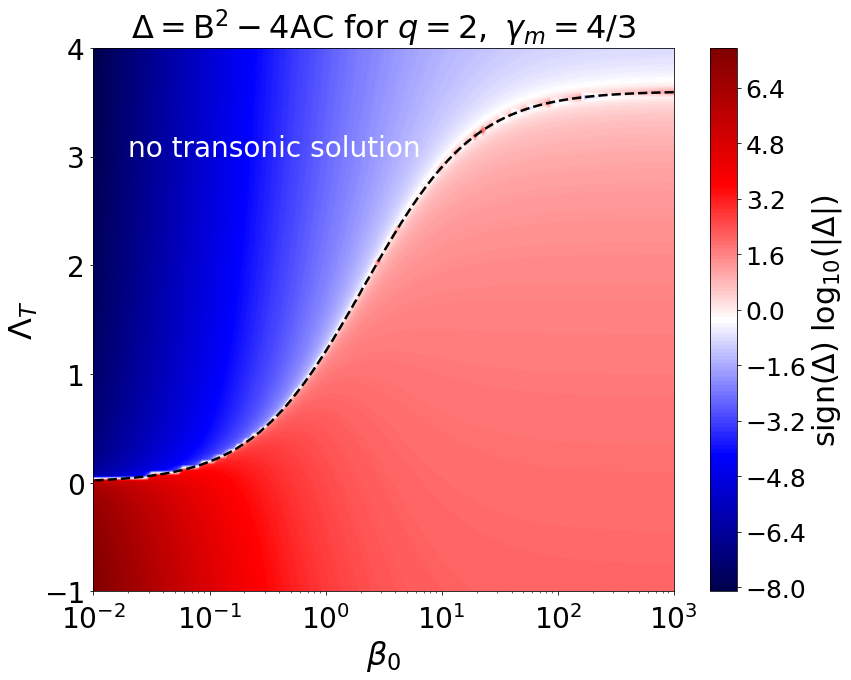}
	\hfill
	\caption [Discriminant in MHD]{Contour plot of the discriminant ($\rm \Delta = B^2-4AC$) in the $\rm \Lambda_T-\beta_0$
	plane for $\rm q=2$ and $\rm \gamma_m=4/3$. The behavior is qualitatively similar for $\rm q=1$ and other $\rm \gamma_m$s. Magnetic fields reduce the parameter space for a transonic solution by providing additional support against cooling-induced inflow. Because the discriminant has orders of magnitude variation in both positive and negative values, the quantity we used for the color is sign($\rm \Delta$)$\log_{10}(|\Delta |)$ which gives visual clarity in depicting the variation of $\rm \Delta$ on both positive and negative sides.}
	\label{fig:discriminant_MHD}
\end{figure}

Like our hydrodynamical cooling flow solution, when including magnetic fields we can also obtain the condition for a transonic solution by expanding the 0/0 form of $\rm d\tilde{v}/d\tilde{r}$ in the wind equation (Eq. \ref{eq:dd_wind_mag}) at the sonic point. In the MHD case, at the sonic point $\rm c_{s0}^2/v_0^2 = (1+\gamma_m/\gamma\beta_0)^{-1}$ and the derivatives there become $\rm \tilde{p}^\prime = \gamma \tilde{v}^\prime + q \gamma_m/\beta_0$ (Eq. \ref{eq:dd_mom_mag}) and $\rm \Lambda^\prime = \Lambda_T([\gamma-1]\tilde v^\prime + q[1+ \gamma_m/\beta_0])$. Following the procedure outlined in section \ref{sec:transonic_condn}, we again obtain a quadratic equation for $\rm \tilde{v}^\prime$ at the sonic point of the form $\rm A\tilde{v}^2 + B \tilde{v} + C=0$, with the coefficients given by
\bse
\begin{align}
 & \rm A  =  \gamma+1 + \frac{\gamma_m}{\gamma \beta_0}(1+\gamma_m)\ , \\
 & \rm B =  q \Big[ 4-\gamma + \Lambda_T (\gamma-1) - \frac{\gamma_m}{\gamma \beta_0} \{ 2\gamma_m - \gamma -4 - (\gamma-1) \Lambda_T \}  \Big]\ , \\
\nonumber
 & \rm C = q \Big[ q(\Lambda_T-2) + 1 - \\ 
 & \hspace{1 cm} \rm \frac{\gamma_m}{\gamma\beta_0} \big\{ q ( 2 + \gamma - ( 1+\gamma+\gamma_m/\beta_0 ) \Lambda_T -\gamma_m ) - 1 \big\} \Big].
\end{align}
\ese
As expected, this expression reduces to the pure hydrodynamical result (Eq. \ref{eq:quadratic}) for $\rm \beta_0 \gg 1$. Figure \ref{fig:discriminant_MHD} shows the parameter space in the $\Lambda_T-\beta_0$ plane for the existence of the transonic cooling flow solution in the presence of magnetic field modelled as an additional polytropic fluid.

\section{Convergence and robustness of time dependent cooling flow profiles}
\label{sec:Appendix_B}

\begin{figure*}
    \centering
\includegraphics[width=0.9\textwidth]{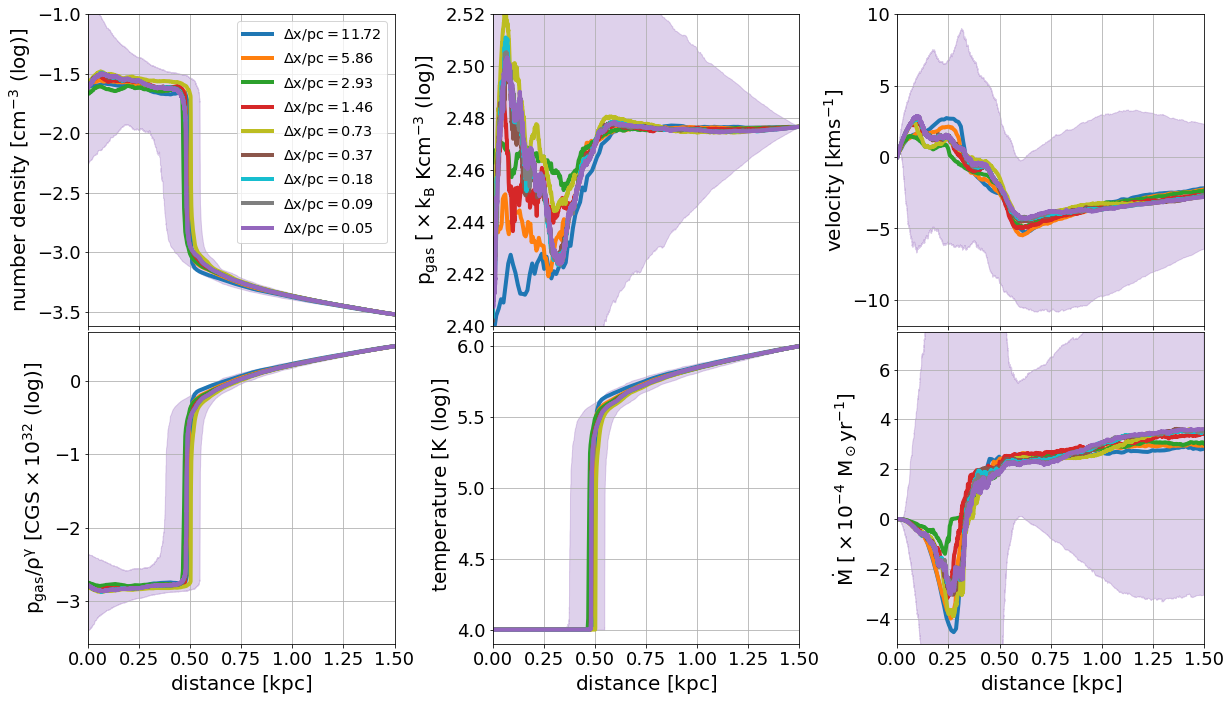}
\caption{The median profiles of number density, pressure, velocity, entropy, temperature and mass inflow rate over all the time snapshots between $\rm 0.7\ t_{cool,hot}$ and $\rm 3\ t_{cool,hot}$ ($\rm t_{cool,hot}$ is the gas cooling time at the outer boundary) generated using the PLUTO code (snapshots are separated by $\rm 2\times 10^{-3}\ t_{cool,hot}$). Each of the individual color represents identical setup run at a different resolution. 
The shaded regions indicate the spread between $\rm 16$ and $\rm 84$ percentile values of the respective quantities at the highest resolution (spreads are similar even for lower resolutions that we do not show). 
The initial and boundary conditions used in these runs are identical to those used in section~\ref{sec:num_sims}. 
}
\label{fig:sphr-pde_convergence}
\end{figure*}

Here we present a convergence study of the time-dependent profiles shown in section~\ref{sec:num_sims}. In Figure~\ref{fig:sphr-pde_convergence} we see numerical convergence of the cooling flow profiles generated by PLUTO in 1D spherical geometry at different numerical resolutions. The median profiles and their spread indicate that gas pressure and velocity are most susceptible to fluctuations. These acoustic fluctuations are generated due to reflections off the very high density gradient at $\sim 500\ {\rm pc}$. 
Our 1D profiles show convergence even at resolutions that don't resolve the cooling length, $\rm c_s t_{cool}$.

We also try different boundary conditions corresponding to different values of $\rm t_{\rm cool}/t_{\rm sc}$ (ratio of cooling time to the sound-crossing time; we vary the boundary density/temperature and the radius of the outer boundary). A smaller value of this ratio gives a larger pressure difference and a higher amplitude of acoustic fluctuations. We find that pressure-driven cooling flows are generated for a range of similar boundary conditions.

\begin{figure*}
    \centering
\includegraphics[width=0.8\textwidth]{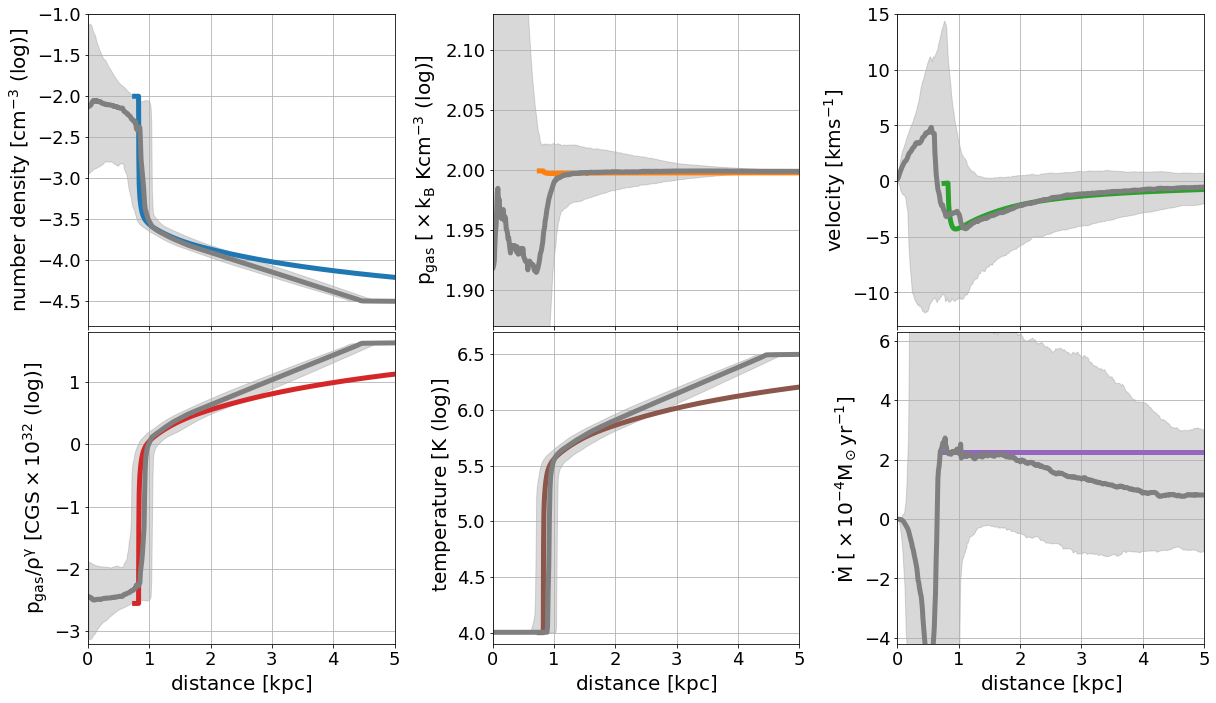}
\caption{The gray lines show the median profiles of number density, pressure, velocity, entropy, temperature and mass inflow rate over all the time snapshots within a time interval of approximately $\rm 1\ Gyr$ generated by the PLUTO code in spherical geometry (q=2) for a radially decreasing initial density. The gray shaded region denotes the spread by showing the $\rm 15$ and $\rm 84$ percentile values. The colored curves indicate the steady cooling flow solution obtained by solving the ODEs (Eqs. \ref{eq:matrix_eqs}; the critical point parameters are $\rm T_0 = 3.3\times 10^{5} K$, $\rm n_0=3 \times 10^{-4} cm^{-3}$ and a Mach number $\rm \mathcal{M}_0 = 0.05$, giving a critical radius of approximately $ 950\  \rm pc$). 
Unlike in section~\ref{sec:num_sims}, we initialize the gas with a log-linear temperature profile varying between $\rm 10^5 K$ at the innermost point and $\rm 3.16\times 10^6 K$ at the outer reservoir, maintaining a constant pressure $\rm p/k_B=100$. 
}
\label{fig:sphr-pde_non-const}
\end{figure*}

We demonstrate the robustness of our steady state model in Figure~\ref{fig:sphr-pde_non-const}. In this setup, we initialize an outward-decreasing isobaric density profile (unlike section~\ref{sec:num_sims}, where initial density/temperature is uniform and outer density/temperature are held fixed). In steady state, the profiles attain steady cooling flow solutions. 
Here, we initialize the gas with a log-linear temperature profile varying between $\rm 10^5 K$ at the innermost grid and $\rm 3.16\times 10^6 K$ at the outer boundary. The density at the outer boundary is fixed to $\rm 3.6\times 10^{-5}\ cm^{-3}$ and pressure is constant at $\rm p/k_B=100$ ${\rm K~cm^{-3}}$. The gas properties at the outermost radius result in a long cooling time, chosen to correspond approximately to the hottest volume filling gas in the TNG50 halo that we analyze (see Figure \ref{fig:dist-evolve}). Since the cooling time at large radii is very long, a steady cooling flow is established within the radius where the cooling time is shorter than the time the system is evolved. The bottom-right panel of Figure \ref{fig:sphr-pde_non-const} shows that the median mass accretion rate is constant only till $\approx 2$ kpc. 
Therefore it is only within this radius, that the steady cooling flow solution matches the PLUTO profiles, indicating the robustness of the cooling flow solutions.

\section{Evolution of CGM temperature and density in TNG50}
\label{sec:Appendix_C}

\begin{figure*}
    \centering
    \includegraphics[width=0.49\textwidth]{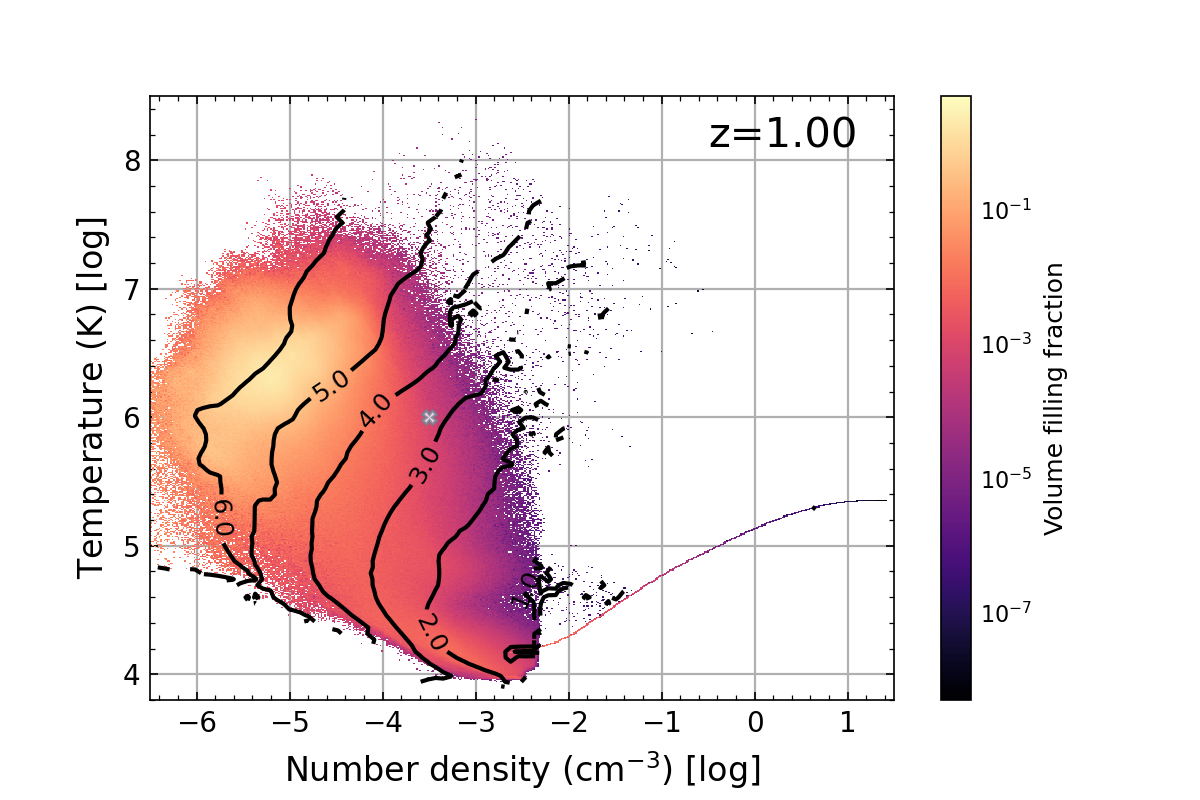}
	\includegraphics[width=0.49\textwidth]{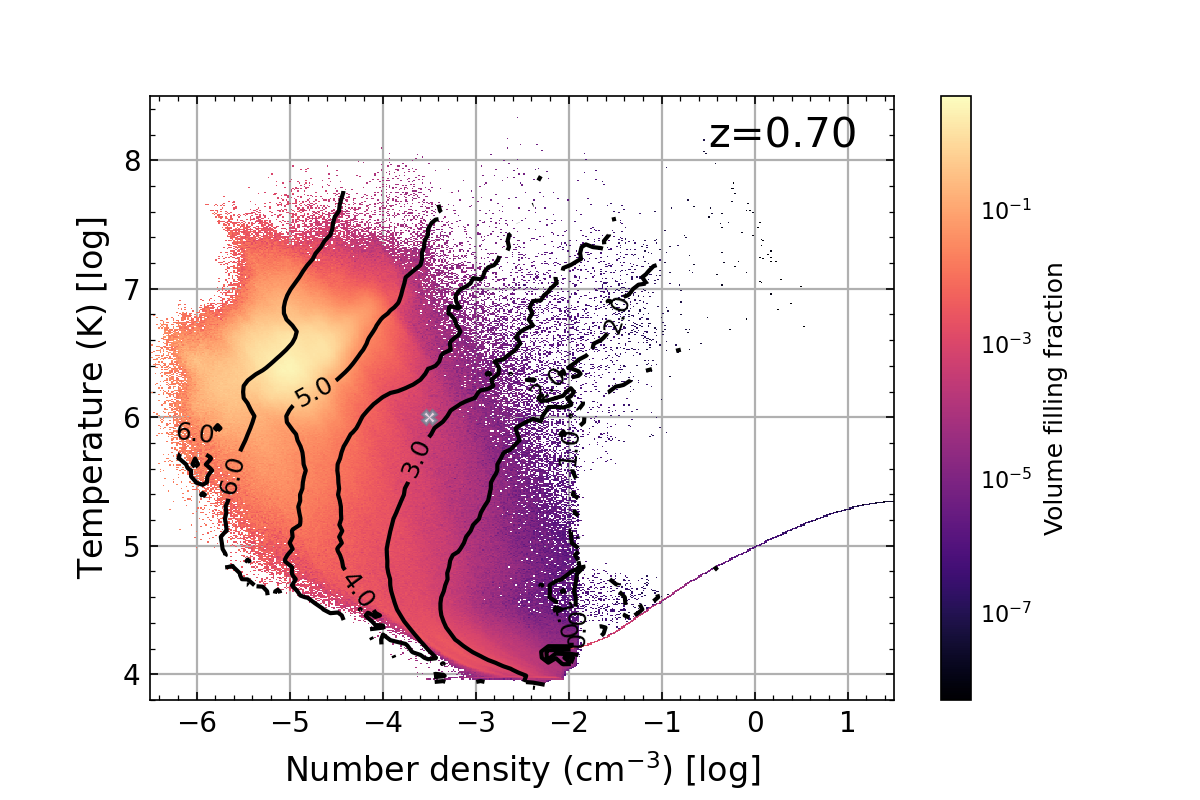}
	\includegraphics[width=0.49\textwidth]{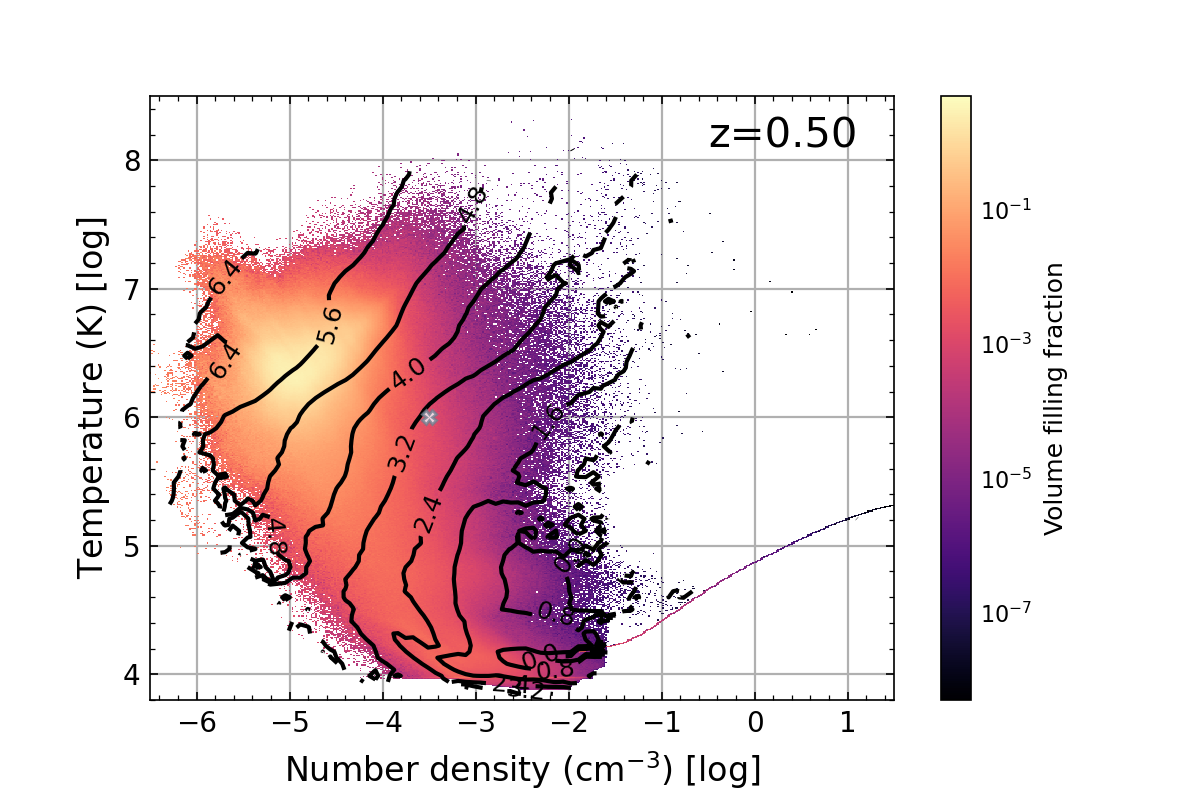}
	\includegraphics[width=0.49\textwidth]{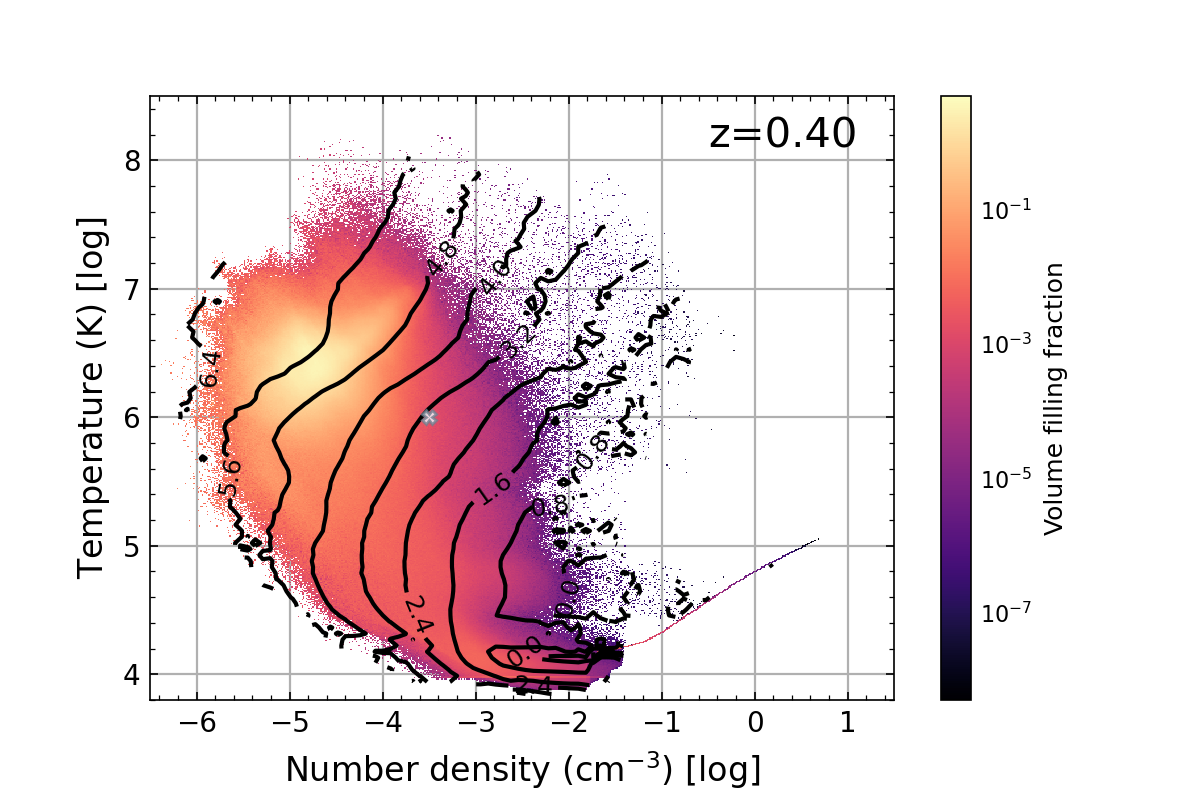}
	\includegraphics[width=0.49\textwidth]{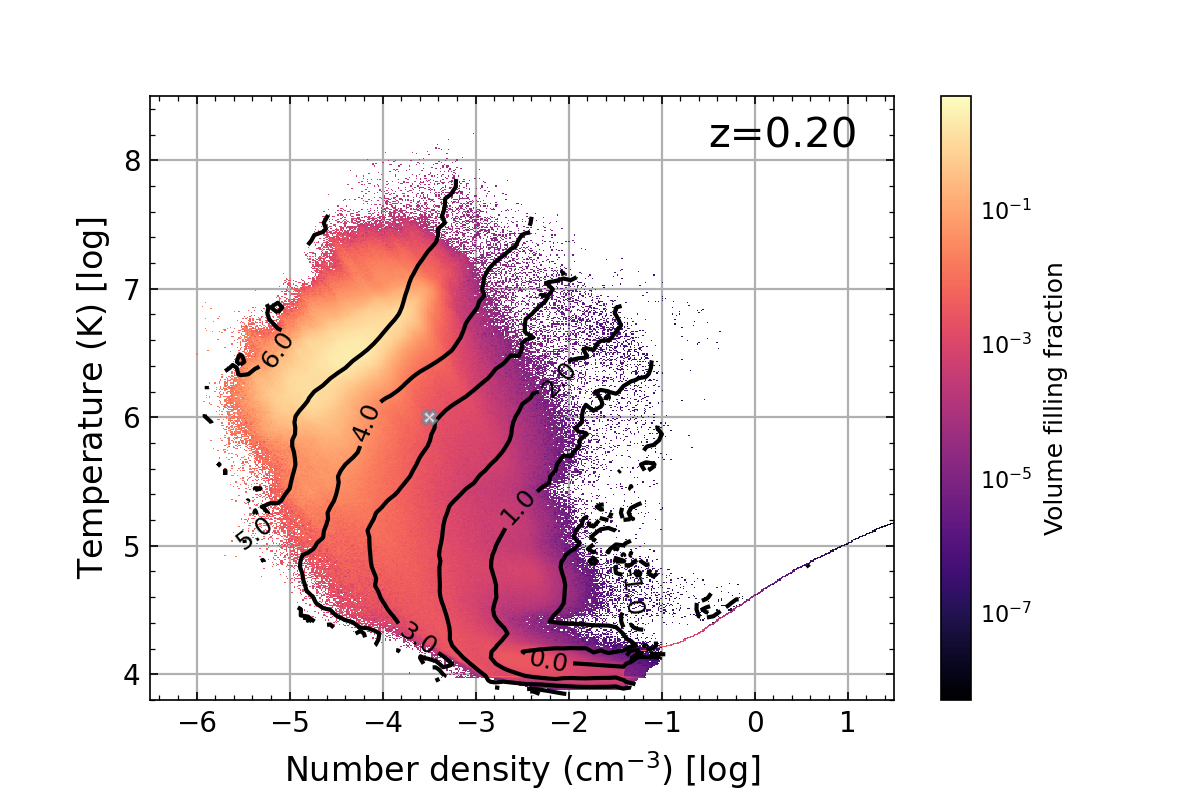}
	\includegraphics[width=0.49\textwidth]{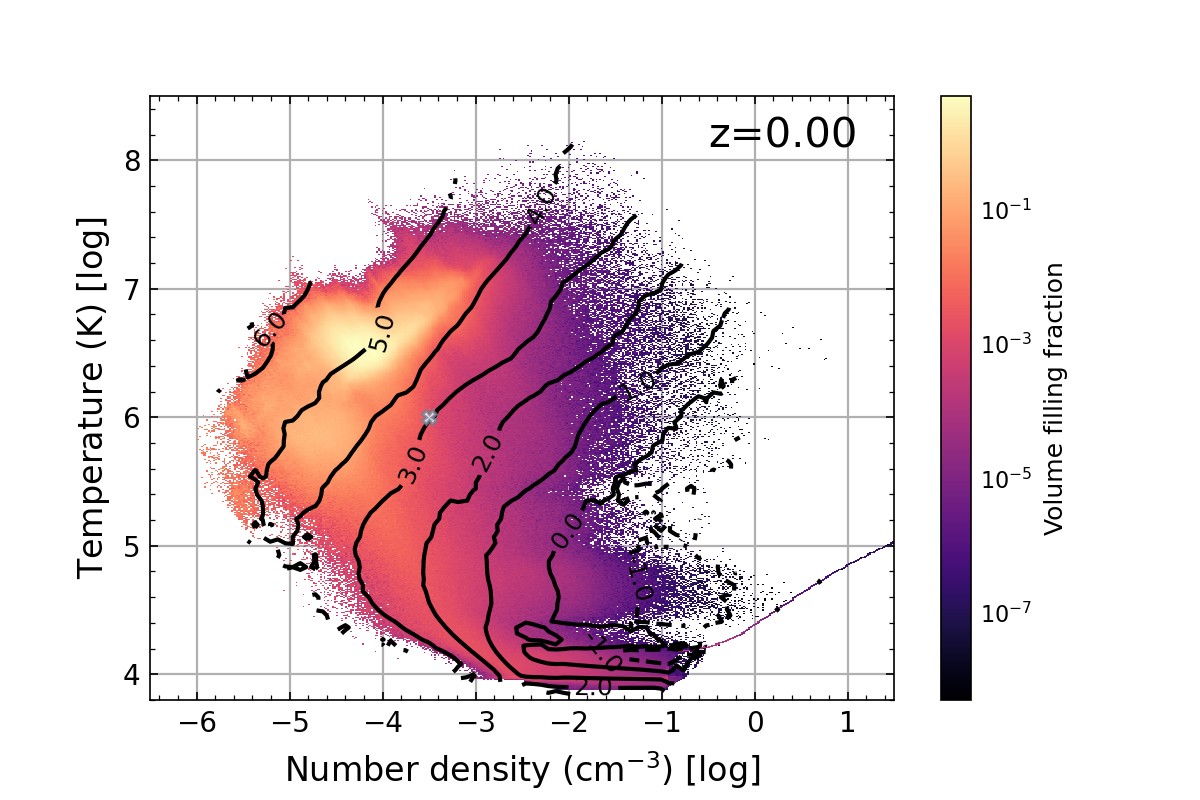}
	\caption{The volume-weighted temperature-density (in physical and not comoving units) phase diagram at different redshifts for our TNG50 halo. The white cross indicates the number density and temperature for our outer boundary in the simulations presented in section~\ref{sec:num_sims}. The black contours show the cooling time of the gas in $\rm log_{10}( Myr)$. The phase distribution reveals that the hot gas is volume-filling and that the hot/intermediate temperature gas is long-lived. Note that the thin feature at high density reflects the two-phase ISM pressurization model of the TNG simulations.}
\label{fig:dist-evolve}
\end{figure*}

Figure~\ref{fig:dist-evolve} shows the distribution of the temperature and density of the halo gas for our fiducial TNG50 halo, followed from $z=0$ to $z=1$. We find that the hot/intermediate temperature phase is volume-filling, and is maintained at approximately fixed temperature and density over timescales exceeding the cooling time of the intermediate phase. We adopt this result to motivate the fixed density/temperature outer boundary conditions for our steady cooling flow setup in section \ref{sec:num_sims}. 

\label{lastpage}
\end{document}